\documentstyle[seceq,preprint]{ptptex}

\newcommand{\lambdabm}{\mbox{\boldmath $\lambda$}}
\newcommand{\sigmabm}{\mbox{\boldmath $\sigma$}}
\newcommand{\xibm}{\mbox{\boldmath $\xi$}}
\newcommand{\fbm}{\mbox{\boldmath $f$}}
\newcommand{\pbm}{\mbox{\boldmath $p$}}
\newcommand{\qbm}{\mbox{\boldmath $q$}}
\newcommand{\rbm}{\mbox{\boldmath $r$}}
\newcommand{\Rbm}{\mbox{\boldmath $R$}}
\makeatletter
\def\lsim{\mathrel{\mathpalette\gl@align<}}
\def\gsim{\mathrel{\mathpalette\gl@align>}}
\def\gl@align#1#2{\lower.6ex\vbox{\baselineskip\z@skip\lineskip\z@
    \ialign{$\m@th#1\hfil##\hfil$\crcr#2\crcr\sim\crcr}}}
\makeatother

\preprintnumber[15cm]{
Submitted to Progress of Theoretical Physics Supplement}

\title{
H-dibaryon
}

\author{
Tsutomu {\sc Sakai}\footnote{E-mail address: tsakai@rcnp.osaka-u.ac.jp},   
Kiyotaka {\sc Shimizu}\footnote{E-mail address:
k-simizu@hoffman.cc.sophia.ac.jp} 
and Koichi {\sc Yazaki}\footnote{E-mail address: 
yazaki@phys.s.u-tokyo.ac.jp}$^{,}$\footnote{Present address: College of Arts 
and Sciences, Tokyo Women's Christian University, Tokyo 167-8585}
}

\inst{
$^{*}$RCNP, Osaka University, Osaka 567-0047
\\
$^{**}$Department of Physics, Sophia University, Tokyo 102-0094
\\
$^{***}$Department of Physics, University of Tokyo, Tokyo 113-0033
}

\recdate{
\today
}

\abst{%
The quark cluster model studies concerning the H-dibaryon are reviewed.
The covered topics are the H-dibaryon itself, the interaction between a nucleon
and an H-dibaryon, and the one between two H-dibaryons.
A related study on the H-dibaryon in nuclear matter is also reviewed and its 
implication to double hypernuclei is discussed.
}

\begin{document}

\input{epsf}

\maketitle

\section{Introduction}

The H-dibaryon is a spin and isospin singlet, positive parity state composed 
of six quarks (uuddss).
It was first proposed by Jaffe in 1977 using MIT bag model as a strongly bound 
state with its mass 81 MeV lighter than the $\Lambda\Lambda$ threshold 
\cite{Jaffe77}.
Being the ground state in the $S=-2$ sector of a $B=2$ system, the H-dibaryon 
is stable against the strong interaction and can decay only via the weak 
interaction.
During two decades, the H-dibaryon has attracted much interest as a 
plausible candidate of exotic states, which are different from the 
hadrons known so far, i.e. mesons $q\bar{q}$ and baryons $q^3$.

Lots of efforts for H-dibaryon hunting have been made although there have
been no conclusive experimental results on the existence of the 
H-dibaryon.\cite{Carroll78}\tocite{Klingenberg99}
Many of the H-dibaryon search experiments have been made with nuclear
reactions, in which the H-dibaryon is expected to be produced in nuclei.

On the theoretical side, many calculations of the H-dibaryon mass and structure
have been performed using various models and
theories.\cite{Jaffe77}\tocite{Zenczykowski87}
Among them, the non-relativistic quark cluster model (QCM), which is 
successful in 
describing the baryon mass spectra and the experimental data of the 
nucleon--nucleon (NN)\cite{Oka81} and nucleon--hyperon 
(NY)\cite{Oka83,Straub88} 
scattering, is applied to the H-dibaryon state.\cite{Oka83}\tocite{Shen99}
Theoretical analyses of the 
production\cite{Badalyan82,Aerts82}\tocite{Kahana99}  
of the H-dibaryon in various processes have been performed mainly by the 
coalescence model.
The weak decay processes of the H-dibaryon are also studied.\footnote{Weak 
decay process of a hypothetical $^1S_0$ bound state of two $\Lambda$'s is also 
discussed by Krivoruchenko and Shchepkin.\cite{Krivoruchenko82}}
Donoghue {\it et al.} calculated the H-dibaryon 
lifetime.\cite{Donoghue86a,Donoghue86b}
For Cygnus X-3 events\cite{Baym85} and ultra-high energy cosmic ray (UHECR) 
events\cite{Kochelev99} beyond the GZK cutoff,\cite{Greisen66,Zatsepin66} 
the H-dibaryon, with its mass below N$\Lambda$ for Cygnus X-3\cite{Baym85} and 
below NN for UHECR,\cite{Kochelev99} has been proposed as a possible 
long-lived, neutral particle which can 
reach the Earth without altering its direction by interstellar magnetic fields 
and losing its energy by the interaction with the cosmic background radiation.
However, the calculated H-dibaryon lifetime is too short to explain the Cygnus 
X-3 events.\cite{Donoghue86a,Donoghue86b}
The H-dibaryon mass is now known to be higher than $\Lambda$N threshold, 
which makes the lifetime even shorter by many orders of magnitude  
and excludes the hypothesis for the UHECR events.

The H-dibaryon may exist in another environment e.g. in a double hypernucleus 
or in some special astrophysical objects.
Double hypernucleus data have important meaning for the existence of the
H-dibaryon in the sense that the binding energy of two $\Lambda$'s is related 
to the lower limit of the H-dibaryon mass.
However, whether $S=-2$ component in a double hypernucleus takes the form of
$\Lambda\Lambda$ is not a trivial problem, and it is possible that a double
hypernucleus is an H-nucleus state.\cite{Sakai95,May98a}
The possibility that H-dibaryon matter may exist in the core of a neutron star 
has also been pointed out.\cite{Tamagaki91}

In this article, we will review the studies of the H-dibaryon using the 
non-relativistic quark cluster model.
In the next section, we will briefly summarize the theoretical and experimental
status of the H-dibaryon.
In \S\ref{sec:H-QCM}, studies on the mass and structure of the H-dibaryon 
employing the quark cluster model will be reviewed.
Though many of them are devoted to the baryon--baryon interaction including the
$S=-2$, $J=T=0$ channel, we will confine ourselves to the H-dibaryon state.
In \S\ref{sec:NH}, a study on the interaction between a nucleon and an
H-dibaryon will be reviewed.
This NH interaction is used for an investigation on the property of the 
H-dibaryon in nuclear matter.
The implication on the double hypernuclei will be also mentioned.
In \S\ref{sec:HH}, we will review a study on the interaction between two
H-dibaryons and the expected properties of H-dibaryon matter are discussed.

\section{H-dibaryon}
\label{sec:H}

In this section, the present status of theoretical and experimental studies on
the H-dibaryon is briefly reviewed.
Double hypernuclei, which have close connection with the H-dibaryon, are also
reviewed.

\subsection{Theoretical status of the H-dibaryon}

\begin{figure}
\epsfxsize=16cm
\centerline{\epsfbox{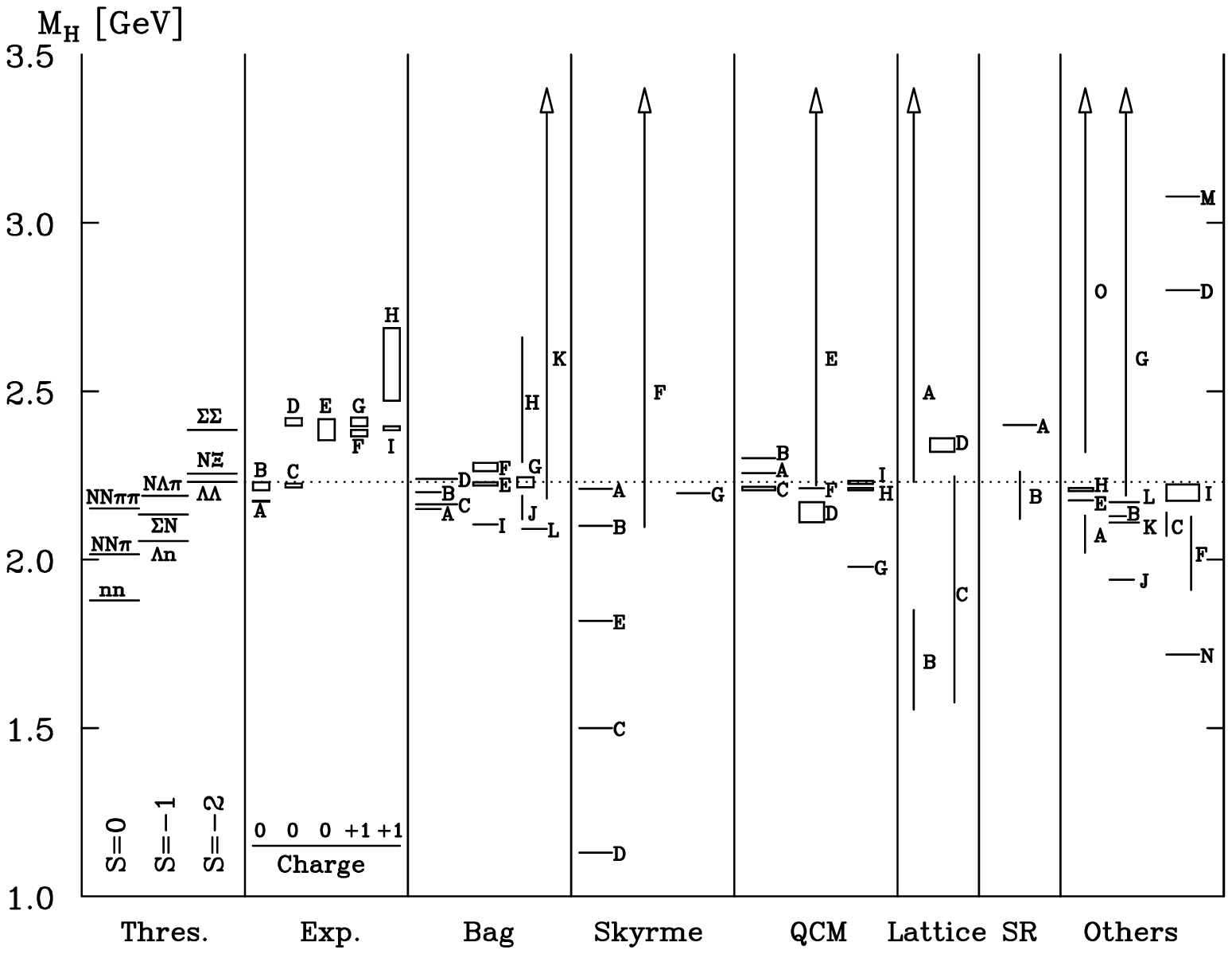}}
\caption{The calculated masses of the H-dibaryon. The dotted line indicates 
the $\Lambda\Lambda$ threshold. Some thresholds\protect{\cite{PDG98}} (Thres.) 
and experimental masses (Exp.) of the H-dibaryon candidates reported so far 
are also shown. 
Vertical lines above the $\Lambda\Lambda$ threshold with upward arrow means 
that the upper limits are not shown in the literature. 
Corresponding references are as follows.
Experiment (Exp.): A:{\protect\citen{Shahbazian88}}, 
B:{\protect\citen{Shahbazian90}}, C:{\protect\citen{Alekseev90}}, 
D--G:{\protect\citen{Shahbazian93}}, H,I:{\protect\citen{Aslanyan99}}. 
The events D--I are claimed to be excited states.
Bag model (Bag): A:{\protect\citen{Jaffe77}},
B:{\protect\citen{Aerts78}}, C:{\protect\citen{Mulders80}}, 
D:{\protect\citen{Liu82}}, E:{\protect\citen{Mulders83}}, 
F:{\protect\citen{Aerts84}}, G:{\protect\citen{Saito84}}, 
H:{\protect\citen{Kondratyuk87}}, I:{\protect\citen{Fleck89}}, 
J:{\protect\citen{Golowich92}}, K:{\protect\citen{Maltman92}}, 
L:{\protect\citen{Dorokhov92}}. 
Skyrme model (Skyrme): A:{\protect\citen{Balachandran84}}, 
B:{\protect\citen{Balachandran85}}, C:{\protect\citen{Jaffe85}}, 
D:{\protect\citen{Yost85}}, E:{\protect\citen{Lee90}}, 
F:{\protect\citen{Scholtz93}}, G:{\protect\citen{Thomas94}}.
The value in Ref.{\protect\citen{Kopeliovich92}} is 3.9$\sim$4.4 GeV.
Non-relativistic quark cluster model (QCM): A:{\protect\citen{Oka83}}, 
B:{\protect\citen{Silvestre87}}, C:{\protect\citen{Straub88}}, 
D:{\protect\citen{Koike90}}, E:{\protect\citen{Takeuchi91,Oka91}}, 
F:{\protect\citen{Nakamoto97}}, G:{\protect\citen{Wolfe97}}, 
H:{\protect\citen{Shimizu99}}, I:{\protect\citen{Shen99}}.
Lattice gauge theory (Lattice): A:{\protect\citen{Mackenzie85}}, 
B:{\protect\citen{Iwasaki88}}, C:{\protect\citen{Yoshie90}}, 
D:{\protect\citen{Negele99}}.
The value in Ref.{\protect\citen{Wetzorke99}} is 2221$\pm$141 MeV for smaller 
lattice and heavier for larger lattice.
QCD sum rule (SR): A:{\protect\citen{Larin86}}, B:{\protect\citen{Kodama94}}.
Others: A:{\protect\citen{Nishikawa91}}, B:{\protect\citen{Pal92}}, 
C:{\protect\citen{Ghosh98}}, D:{\protect\citen{Carlson91}}, 
E:{\protect\citen{Rosner86}}, F:{\protect\citen{Goldman87}}, 
G:{\protect\citen{Fleck89}}, H:{\protect\citen{Aizawa91}}, 
I:{\protect\citen{Wang95}}, J:{\protect\citen{Kondratyuk89}}, 
K:{\protect\citen{Diakonov89}}, L:{\protect\citen{Goldman98}}, 
M:{\protect\citen{Stancu98}}, N:{\protect\citen{Kochelev99}}, 
O:{\protect\citen{Lichtenberg97}}.
Here we cite the values in the references as they are.
Some of them are better to be compared with the mass of two $\Lambda$'s 
calculated in the same framework.
When the cited value is the binding energy, it is subtracted from the 
experimental value of the $\Lambda\Lambda$ threshold (2231 MeV).
For further details please see each reference. 
}
\label{fig:hmass}
\end{figure}

Since Jaffe's prediction,\cite{Jaffe77} many theoretical calculations have 
been made to predict the mass of the
H-dibaryon, employing various QCD-inspired models (bag 
model\cite{Aerts78}\tocite{Dorokhov92}, nonrelativistic quark cluster 
model\cite{Oka83}\tocite{Shen99}, Skyrme
model\cite{Balachandran84}\tocite{Thomas94}, and so
on\cite{Rosner86}\tocite{Gignoux87}), QCD sum rule\cite{Larin86,Kodama94} 
and lattice QCD\cite{Mackenzie85}\tocite{Wetzorke99}.
Many of them predict the bound state.
However, the results of calculations spread over wide range as shown in 
Fig.~\ref{fig:hmass}.
It makes clear contrast to the success in reproducing the mass spectrum 
of mesons and baryons using the above methods.

This situation indicates an importance of the H-dibaryon as a touchstone 
of the predictive power of models and theories, and, on the other hand, 
experimental studies on the existence of the H-dibaryon are highly 
encouraged and will give a new insight into the dynamics of multiquark 
systems.

The basic mechanism which is expected to give large attractive force 
between quarks in the H-dibaryon is the color magnetic interaction (CMI),
which is proportional to 
$\sigmabm_i \cdot \sigmabm_j \lambdabm_i \cdot \lambdabm_j$, 
where $\sigmabm_i$ is the Pauli matrix for the spin SU(2) group and 
$\lambdabm_i$ is the Gell-Mann matrix for the color SU(3) group of the $i$th 
quark.
In the flavor SU(3) symmetric limit, the expectation value of the sum of 
the operator $\sigmabm_i \cdot \sigmabm_j \lambdabm_i \cdot \lambdabm_j$ 
for all pairs of quarks in a color singlet $n$-quark system with negative 
sign is
\begin{equation}
\Theta \equiv - \langle \sum_{i<j} \sigmabm_i \cdot \sigmabm_j 
\lambdabm_i \cdot \lambdabm_j \rangle = n(n-10) + \frac{4}{3} J(J+1) + 
\left\langle \left( \sum_{i} \fbm_i \right)^2 \right\rangle ,
\label{CMI-op}
\end{equation}
where $J$ is the spin of the $n$-quark system and 
\begin{equation}
\left\langle \left( \sum \fbm_i \right)^2 \right\rangle = 
\frac{4}{3} \left( \lambda^2 + \lambda \mu + \mu^2 \right) + 4(\lambda + \mu) .
\end{equation}
Here, $(\lambda, \mu)$ specifies the irreducible representation of the flavor 
SU(3) group.
$\Theta$ favors the spin-flavor symmetric state.
In fact, the expectation value of the Casimir operator 
$\left\langle \left( \sum \fbm_i \right)^2 \right\rangle$ is 12 for 
$[21]_{\rm f} = (1, 1)$ state\cite{Shimizu89} and 0 for 
the flavor-singlet state like the H-dibaryon, so that $2\Theta = -16$ for 
two $\Lambda$'s and $\Theta=-24$ for the H-dibaryon.
This is the main reason why the H-dibaryon is likely to be a bound state.
In terms of baryon configuration, the flavor SU(3) symmetric state 
[222]$_{\rm f}$ can be written as
\begin{equation}
\mid {\rm H} \rangle =  \sqrt{\frac{1}{8}} \mid \Lambda\Lambda \rangle 
                      + \sqrt{\frac{4}{8}} \mid {\rm N}\Xi \rangle
                      - \sqrt{\frac{3}{8}} \mid \Sigma\Sigma \rangle .
\label{HSU3wf}
\end{equation}
Recent QCD sum rule calculation proposed another possibility of the mass 
spectrum of the ``H-dibaryons"\footnote{The term H-dibaryon is used for the 
flavor-singlet H-dibaryon state in this paper. For other multiplets of uuddss 
color-singlet 
states,\cite{Jaffe77,Aerts78,Kondratyuk89,Bickerstaff81}\tocite{Paganis97,Paganis99} 
we refer the flavor multiplets they belong.
The unitary transformation coefficients between each multiplets and two-baryon 
configurations are shown in Table~3 of Ref.~\protect\citen{Oka91}.
}, 
i.e. 27plet $[42]_{\rm f}$ and octet $[321]_{\rm f}$ H-dibaryons 
are lighter than the singlet H-dibaryon although they are resonances 
between $\Lambda\Lambda$ and $N\Xi$ thresholds in their 
calculation\cite{Paganis97}.

\subsection{Experimental status of the H-dibaryon}
\label{subsec:Hexp}

Experimental searches of H-dibaryon have been performed elaborately.
However, there have been no conclusive results on the existence of the 
H-dibaryon.
The H-dibaryon searches have been made by several methods and in the wide 
range of its expected mass and lifetime.
The H-dibaryon may be produced via $(K^-, K^+)$ reaction, $\Xi^-$-capture, 
heavy ion 
collision, $\bar{p}$-nucleus annihilation reaction and so on, and then the 
fragments produced in the decay of the H-dibaryon are trucked, e.g. for the 
H-dibaryon below $\Lambda\Lambda$ threshold, 
${\rm H} \rightarrow \Sigma^- p$, ${\rm H} \rightarrow \Sigma^0 n$, 
${\rm H} \rightarrow \Lambda n$, ${\rm H} \rightarrow p \pi^- \Lambda$, and 
so on.
We do not step into the details of each experiment here.
Instead, we list in Table~\ref{tab:experiment} the experiments with their 
key reaction processes, i.e. how to produce the H-dibaryon, the decay 
processes to detect, or the process like $pp \rightarrow K^+K^+X$ in which 
the invariant mass analysis is made.
Many of them are still in progress.
See also the review paper by Paul\cite{Paul91} including ealier 
experiments\cite{Beilliere72,Wilquet75,Goyal80} not shown in 
Table~\ref{tab:experiment} and recent reviews by Ashery\cite{Ashery98} and 
Klingenberg\cite{Klingenberg99}.

\begin{table}[t]
\caption{Experimental searches for the H-dibaryon.
For some of the experiments, we show in the third column to what range of 
the H-dibaryon mass that experiment is sensitive.
For KEK E224, $(pp)$ and $(p)$ mean a proton pair and a proton in 
${}^{12}{\rm C}$, respectively.
$B_{\rm H}\equiv 2M_{\Lambda}-M_{\rm H}$.
}
\label{tab:experiment}
\begin{center}
\begin{tabular}{lll} \hline \hline
Collaboration & reaction process (production/decay) & sensitive mass range \\ 
\hline
BNL E703\protect{\cite{Carroll78}} & $p+p \rightarrow K^++K^++X$ & 
$M_{\rm H}=2.0 \sim 2.5$ GeV \\
BNL E810\protect{\cite{Longacre95,Longacre98,Bassalleck97}} & 
Si+Pb collision / ${\rm H} \rightarrow \Sigma^- p , \Lambda p \pi^-$ & \\
BNL E813 & $K^-+p \rightarrow K^++\Xi^-$, 
$(\Xi^-{\rm d})_{\rm atom} \rightarrow {\rm H}+n$ & 
$-15 < B_{\rm H} <80 $ MeV \\
\hspace*{3mm}\protect{\cite{Franklin86,Iijima94,Barnes92,Ramsay97,Quinn93,Bassalleck95,Bassalleck97,Bassalleck98}} 
& & \\
BNL E830\protect{\cite{Chrien98}} & $K^-+{}^3{\rm He} \rightarrow 
K^++{\rm H}+n$ & \\
BNL E836 
& $K^-+{}^3{\rm He} \rightarrow K^++{\rm H}+n$ & $B_{\rm H}=50 \sim 380$ MeV \\
\hspace*{3mm}\protect{\cite{Barnes92,Ramsay97,Quinn93,Stotzer97,Bassalleck95,Bassalleck97,Bassalleck98}}
 & $K^-+{}^6{\rm Li} \rightarrow K^++{\rm H}+X$ & \\
BNL E864\protect{\cite{Bassalleck97,Chrien98}} & Au+Pb collision & \\
BNL E885\protect{\cite{Quinn93,Yamamoto98,May98b,Bassalleck97}} & 
$K^-+(p) \rightarrow K^++\Xi^-$, & \\
 & $(\Xi^-A)_{\rm atom} \rightarrow {\rm H} +X$ & \\
 & $K^-+A \rightarrow K^++X+{\rm H}$ & \\
BNL E886\protect{\cite{Rusek95,Bassalleck97}} & Au+Pt collision & \\
BNL E888 & $p + A \rightarrow {\rm H} + X$ / ${\rm H} \rightarrow \Lambda n$ 
or $\Sigma^0 n$, & \\
\hspace*{3mm}\protect{\cite{Belz95,Belz96a,Belz96b,Bassalleck97,Bassalleck98}} 
& \hspace{3mm} H+$A \rightarrow \Lambda +\Lambda +A$ & 
$M_{\rm H} < 2150$ MeV \\
BNL E896\protect{\cite{Crawford98,Bassalleck97,Chrien98}} & Au+Au collision / 
${\rm H}\rightarrow \Sigma^-p \rightarrow n \pi^-p$, & \\
 & 
${\rm H}\rightarrow \Lambda p\pi^- \rightarrow p \pi^- p \pi^-$, 
${\rm H}\rightarrow \Lambda n \rightarrow p \pi^-n$ 
& \\
BNL E910\protect{\cite{Chemakin98}} & $p$+$A$ / 
${\rm H}\rightarrow \Lambda p \pi^-$, ${\rm H}\rightarrow \Sigma^-p$ & \\
BNL STAR\protect{\cite{Coffin97,Paganis99}} & 
Au+Au collision & \\
KEK E176\protect{\cite{Aoki90,Imai91,Imai92,Imai93}} & 
   $K^-+(pp) \rightarrow K^++{\rm H}$ & \\
 & $K^-+p \rightarrow K^++\Xi^-$, $\Xi^-+(p) \rightarrow {\rm H}$ & \\
KEK E224\protect{\cite{Ahn92,Ahn94,Ahn96,Ahn98,Ahn98b,Imai93}} 
 & $K^-+(pp) \rightarrow K^++{\rm H}$ & \\
 & $K^-+(p) \rightarrow K^++\Xi^-$, $\Xi^-+(p) \rightarrow {\rm H}$ & \\
KEK E248\protect{\cite{Nakazawa98}} & $p+p \rightarrow K^++K^++X$ & \\
Fermilab E791\protect{\cite{Ashery96}} 
 & ${\rm H} \rightarrow p+\pi^-+\Lambda$, $\Lambda \rightarrow p+\pi^-$, & \\
 & ${\rm H} \rightarrow \Lambda+\Lambda \rightarrow p+\pi^-+p+\pi^-$ & \\
Fermilab KTeV Collab. 
 & p+A / ${\rm H} \rightarrow p+\pi^-+\Lambda$ & $M_{\rm H} = 2194$ \\
\hspace*{7mm}\protect{\cite{Alavi-Harati99}} & & 
\hspace*{6ex} $\sim 2231$ MeV \\
Shahbazian {\it et 
al.}\protect{\cite{Shahbazian88,Shahbazian90,Shahbazian93,Shahbazian94,Aslanyan99}} & 
$p+{}^{12}{\rm C} \rightarrow {\rm H} ({\rm H}^+) +X$ / & \\
 & \hspace{3mm} ${\rm H} \rightarrow \Sigma^- + p$, 
$\Sigma^- \rightarrow \pi^- n$ & \\
 & \hspace{3mm} ${\rm H}^+ \rightarrow p+\pi^0+\Lambda$, 
$\Lambda \rightarrow p+\pi^-$ & \\
 & \hspace{3mm} ${\rm H}^+ \rightarrow p+\Lambda$, 
$\Lambda \rightarrow p+\pi^-$ & \\
Alekseev {\it et al.}\protect{\cite{Alekseev90}} & 
$n+A \rightarrow {\rm H}+X$ / ${\rm H} \rightarrow p\pi^-\Lambda$, $\Lambda
\rightarrow p \pi^-$ & \\
DIANA Collab.\protect{\cite{Barmin93,Barmin96}} & 
$\bar{p} + {\rm Xe} \rightarrow K^+{\rm H}X$, $K^+K^+{\rm H}X$ / & \\
 & \hspace{3mm} ${\rm H} \rightarrow \Sigma^-+p$ & \\
Condo {\it et al.}\protect{\cite{Condo84}} & 
$\bar{p}+A \rightarrow {\rm H}+X$ / ${\rm H} \rightarrow \Sigma^- + p$ & \\
Ejiri {\it et al.}\protect{\cite{Ejiri89}} & 
${\rm d} \rightarrow {\rm H}+\beta +\nu$, 
${}^{10}{\rm Be} \rightarrow {}^{8}{\rm Be}+{\rm H}$, 
& $M_{\rm H} < 1875.1$ MeV \\
 & ${}^{72}{\rm Ge} \rightarrow {}^{70}{\rm Ge}+{\rm H}+\gamma$, 
${}^{127}{\rm I} \rightarrow {}^{125}{\rm I}+{\rm H}+\gamma$, & \\
 & ${}^{127}{\rm I} \rightarrow {}^{125}{\rm Te}+{\rm H}+\beta^++\nu$ & \\
CERN NA49\protect{\cite{Mitchell95}} & 
Pb+Pb collision / ${\rm H} \rightarrow \Sigma^- p, \, \Lambda p \pi$  & \\
CERN WA89\protect{\cite{Paul94}} & 
$\Sigma^-+A \rightarrow X+{\rm H}$ / ${\rm H} \rightarrow \Lambda\Lambda, 
{\rm N}\Xi$, & \\
 & \qquad \qquad \qquad \quad 
${\rm H} \rightarrow \Lambda p \pi^-, \Sigma^- p, \Sigma^0 n, \Lambda n$ & \\
CERN WA97\protect{\cite{Jacholkowski99}} & Pb+Pb collision & \\
CERN ALICE\protect{\cite{Coffin97}} & Pb+Pb collision & \\
CERN OPAL\protect{\cite{OPAL96}} & $Z^0$ decay & \\
\hline
\end{tabular}
\end{center}
\end{table}

Most of the groups have reported that no evidence was found for the 
H-dibaryon. 
However, Shahbazian et
al.\cite{Shahbazian88,Shahbazian90,Shahbazian93,Shahbazian94,Aslanyan99} 
reported some H-dibaryon candidates including H-dibaryon excited states.
Alekseev {\it et al.}\cite{Alekseev90} also reported two candidate events of 
the H-dibaryon with $M_{\rm H}=2217.1\pm 7.1$ MeV and $2224.3\pm 8.4$ MeV.
(See Exp. part in Fig.~\ref{fig:hmass}. F, G, H and I are positively charged, 
and claimed to be a member of isotriplet together with neutral particle 
events, D and E.)
These events, however, are controversial and more careful background 
event analysis and further experimental supports are needed.
Anyway, few events with different masses cannot lead to any conclusion.
Recently, the enhancement of the $\Lambda\Lambda$ invariant mass near 
threshold was observed in $^{12}$C$(K^-,K^+\Lambda\Lambda)$ reaction (KEK 
E224).\cite{Ahn98b}
However, according to the analysis by Ohnishi et al.,\cite{Ohnishi99} this 
enhancement can be reproduced by the attractive $\Lambda\Lambda$ final state 
interaction with the possible scattering length corresponding to the bound 
or unbound $\Lambda\Lambda$.
At present, whether this enhancement means a $\Lambda\Lambda$ resonance state 
or not is unclear.

Some of the H-dibaryon production processes are analyzed with theoretical 
models, and the H-dibaryon formation cross sections are calculated.
In Table~\ref{tab:Hproduction}, such analyses are listed.
\begin{table}[thb]
\caption{H-dibaryon production processes analyzed theoretically. $^{\ast)}$ 
Formation of a bound state of two $\Lambda$'s as a doorway to the H-dibaryon 
state.}
\label{tab:Hproduction}
\begin{center}
\begin{tabular}{lcll} \hline \hline
Author/Year & Ref. & \multicolumn{2}{l}{process} \\ \hline 
Badalyan {\it et al.} '82 & {\protect\citen{Badalyan82}} & 
\multicolumn{2}{l}{$pp \rightarrow K^+ K^+ {\rm H}$} \\
Aerts \& Dover '82,'83 & {\protect\citen{Aerts82,Aerts83}} & 
\multicolumn{2}{l}{$K^- + ^3{\rm He} \rightarrow K^+ + n + {\rm H}$} \\
Aerts \& Dover '84 & {\protect\citen{Aerts84b}} & 
$K^- + p \rightarrow K^+ + \Xi^-$, \quad &
$(\Xi^-p)_{\rm atom} \rightarrow {\rm H} + \gamma$\\
 & & & $(\Xi^-{\rm d})_{\rm atom} \rightarrow {\rm H} + n$\\
 & & & $(\Xi^-{}^4{\rm He})_{\rm atom} \rightarrow {\rm H} + {\rm t}$
\\
Dover {\it et al.} '89 & {\protect\citen{Dover89}} & 
\multicolumn{2}{l}{Si+Au collision} \\
Sano {\it et al.} '89 & {\protect\citen{Sano89}} & 
\multicolumn{2}{l}{Ne+Ne, $p$+Ne collision} \\
Kishimoto '89 & {\protect\citen{Kishimoto89}} & 
\multicolumn{2}{l}{
$^{\rm A}_{\Lambda}{\rm N} \rightarrow ^{\rm A-2}{\rm N} + {\rm H} + \pi^+$} \\
Dover {\it et al.} '91 & {\protect\citen{Dover91a}} & 
\multicolumn{2}{l}{high-energy nuclear collision} \\
Moinester {\it et al.} '92 & {\protect\citen{Moinester92}} & 
     \multicolumn{2}{l}{
       $\Xi^- + p \rightarrow \pi^0 + {\rm H} , \rho^0 + {\rm H}$} \\
 & & \multicolumn{2}{l}{
       $\Sigma^- + p \rightarrow K^0 + {\rm H} , K^{0\ast} + {\rm H}$} \\
 & & \multicolumn{2}{l}{
       $\Lambda + p \rightarrow K^+ + {\rm H} , K^{\ast +} + {\rm H}$} \\
 & & \multicolumn{2}{l}{
       $\Xi^- + p \rightarrow {\rm H} + \bar{p} + X$} \\
Aizawa \& Hirata '92 & {\protect\citen{Aizawa92}} & 
\multicolumn{2}{l}{$K^- + ^3{\rm He} \rightarrow K^+ + n + {\rm H}$} \\
Rotondo '93 & {\protect\citen{Rotondo93}} & 
\multicolumn{2}{l}{$p$-$A$ collision} \\
Baltz {\it et al.} '94 $^{\ast)}$ & {\protect\citen{Baltz94}} & 
\multicolumn{2}{l}{Au+Au $\rightarrow$ $(\Lambda\Lambda)_{\rm b}+X$} 
\\
Cole {\it et al.} '95 & {\protect\citen{Cole95}} & 
\multicolumn{2}{l}{$p$-$A$ collision} \\
Cousins \& Klein '97 & {\protect\citen{Cousins97}} & 
\multicolumn{2}{l}{$p$-Pt collision} \\
Batty {\it et al.} '99 & {\protect\citen{Batty99}} & 
\multicolumn{2}{l}{$(\Xi^-{\rm d})_{\rm atom} \rightarrow {\rm H} + n$} \\
Kahana '99 & {\protect\citen{Kahana99}} & 
\multicolumn{2}{l}{Au+Au collision} \\
\hline
\end{tabular}
\end{center}
\end{table}
Some of them have been very helpful to experiments.
For instance, the analyses of the H-dibaryon formation processes of BNL E813 
and E836 are given in the papers by Aerts and
Dover,\cite{Aerts82,Aerts83,Aerts84b} and have given guidances to the 
experiments through the direct comparison between theoretical prediction 
and experimental data.
The relation between the lifetime and the mass of the H-dibaryon calculated 
by Donoghue {\it et al.}\cite{Donoghue86a,Donoghue86b} also has played 
important roles in setting the searching mass and lifetime region in 
experiments.
However, as May pointed out,\cite{May98a} this relationship may depend 
critically on the wave function of the H-dibaryon.

\subsection{Double hypernuclei}

The double hypernucleus sheds light on the problem of the existence of the
H-dibaryon from other aspects.
Consider a double hypernucleus which is formed by the fusion of a 
nonstrange nucleus and two $\Lambda$'s with its binding energy
$B_{\Lambda\Lambda}\equiv M(^{A-2}Z)+2M_{\Lambda}-M(^{A}_{\Lambda\Lambda}Z)$.
If the mass of the double hypernucleus is heavier than the sum of the 
masses of the H-dibaryon and the original nucleus, then the double 
hypernucleus can decay strongly into the H-dibaryon and the original nucleus.
Therefore, the existence of a double hypernucleus means that the mass of 
the H-dibaryon should be heavier than the mass of the two $\Lambda$'s 
minus the binding energy: $M_{\rm H} > 2 M_{\Lambda} - B_{\Lambda\Lambda}$.
There have been reported a few events of double hypernuclei.
Old nuclear emulsion experiments reported 
$^{10}_{\Lambda\Lambda}$Be\cite{Danysz63} and 
$^{6}_{\Lambda\Lambda}$He.\cite{Prowse66}
The former was reanalyzed by Dalitz {\it et al.}\cite{Dalitz89}
An emulsion-counter hybrid experiment (KEK E176) reported that an 
event\cite{Aoki91b} can be interpreted as 
$^{10}_{\Lambda\Lambda}$Be\cite{Aoki91b,Yamamoto91} or 
$^{13}_{\Lambda\Lambda}$B.\cite{Aoki91b,Yamamoto91,Dover91b}
Since then several candidates of double hypernuclei have been reported 
and the analyses are under way.\cite{Tanaka95,Nakazawa95}
Many projects for double hypernucleus hunting are going 
on.\cite{Quinn93,May98b,Chrien98,Bassalleck98,Nakazawa98,Fukuda96,Franklin95}\tocite{May98a}
In Table~\ref{tab:doublehyper}, we list the double hypernuclei reported so 
far with their binding energy of two $\Lambda$'s in the double hypernuclei 
$B_{\Lambda\Lambda}$ and the $\Lambda\Lambda$ interaction energy defined by 
$\Delta B_{\Lambda\Lambda}(^A_{\Lambda\Lambda}Z)\equiv 
B_{\Lambda\Lambda}(^A_{\Lambda\Lambda}Z)-2B_{\Lambda}(^{A-1}_{\Lambda}Z)$, 
where $B_{\Lambda}(^A_{\Lambda}Z)\equiv
M(^{A-1}Z)+M_{\Lambda}-M(^A_{\Lambda}Z)$.
\begin{table}
\caption{Reported double hypernuclear events are listed with their two
$\Lambda$ binding energy $B_{\Lambda\Lambda}$ and their $\Lambda\Lambda$ 
interaction energy $\Delta B_{\Lambda\Lambda}$.
$\Delta B_{\Lambda\Lambda}$ for the event reported by Aoki {\it et al.} is 
cited from Refs.{\protect\citen{Yamamoto91,Dover91b}}.}
\label{tab:doublehyper}
\begin{center}
\begin{tabular}{lllrr} \hline \hline
Year & Authors & Nuclide & $B_{\Lambda\Lambda}$ (MeV) & 
$\Delta B_{\Lambda\Lambda}$ (MeV) \\
\hline
1963 & Danysz {\it et al.}\cite{Danysz63} & ${}^{10}_{\Lambda\Lambda}$Be & 
$17.7\pm 0.4$ & $4.3\pm 0.4$ \\
1966 & Prowse {\it et al.}\cite{Prowse66} & ${}^{6}_{\Lambda\Lambda}$He & 
$10.9\pm 0.5$ & $4.7\pm 1.0$ \\
1991 & Aoki {\it et al.}\cite{Aoki91b} & ${}^{10}_{\Lambda\Lambda}$Be & 
$8.5\pm 0.7$ & $-4.9\pm 0.7$ \\
 & & or & & \\
 & & ${}^{13}_{\Lambda\Lambda}$B & $27.6\pm 0.7$ & $4.9\pm 0.7$ \\
\hline 
\end{tabular}
\end{center}
\end{table}

After all, only the H-dibaryon with the small binding energy less than a few 
tens MeV survives in the present status.

\section{H-dibaryon in the quark cluster model}
\label{sec:H-QCM}

In this section, we will review the work on the H-dibaryon in the
non-relativistic quark cluster models (QCMs).
These investigations have been done in relation to $S=-2$, $J^{\pi}=0^+$, 
$T=0$ channel baryon--baryon interaction.
The non-relativistic quark cluster models are successful in reproducing 
baryon mass spectrum and NN scattering phase shifts.
Although the NY scattering data are not enough to make phase shift 
analyses, QCMs are also successful in reproducing the cross sections.
We will not go further into the baryon spectrum and baryon--baryon 
interaction in QCM, and leave them to the other reviews in this supplement and 
a previous review\cite{Shimizu89} by Shimizu.
We only mention here that the results for baryon--baryon interaction 
by the QCMs agree well not only qualitatively but also quantitatively.
In QCM, the internal wave function of a baryon is represented by a shell 
model wave function (usually the one of the harmonic oscillator), which 
is assumed to be known.
Then, the total wave function of a baryon--baryon system is constructed by the 
antisymmetrized product of two baryons supplemented by the wave function of 
the relative motion between two baryons.
The variational principle for energy (bound states)\cite{Wheeler37,Horiuchi77} 
or S-matrix (scattering states)\cite{Kamimura77} leads to the resonating group 
method (RGM) equation, by which the relative motion wave function is solved.
The versions of QCMs are different in the way of reflecting the flavor
symmetry breaking to the wave function, interaction between quarks, the 
treatment for meson exchange contributions, or the treatment of quark 
confinement.

The wave functions for the flavor SU(3) octet baryons with spin 
$\frac{1}{2}$ (N, $\Sigma$, $\Xi$, $\Lambda$) are given by
\begin{equation}
\phi (q^3) = \varphi (\xibm_1, \xibm_2) {\cal F}([21]_f) 
{\cal S}([21]_{\sigma}) {\cal C}([111]_c) ,
\end{equation}
where [21]$_f$, [21]$_{\sigma}$ and [111]$_c$ denote the irreducible 
representations of the flavor, spin and color symmetry, respectively.
For the low-lying baryons, the orbital part is assumed to have [3] 
symmetry.
Most of the calculations assume the flavor SU(3) symmetry in the wave 
functions, although the symmetry breaking is taken into account in the 
hamiltonian.
A few versions of QCMs take account of the mass difference between the 
non-strange and strange quarks also in the wave function.
The orbital part of the single quark wave function in the baryon 
is taken to be Gaussian, i.e. 
\begin{equation}
\varphi (\rbm) = (\pi b^2)^{-3/4} e^{-\frac{r^2}{2b^2}} ,
\end{equation}
where $b$ is the size parameter of the harmonic oscillator.
The flavor SU(3) symmetry breaking in $\varphi (\rbm)$ is taken into 
account by the replacement:\cite{Straub88,Fleck89,Aizawa91}
\begin{equation}
\frac{1}{b_{\rm s}^2} = \frac{m_{\rm s}}{m_{\rm u,d}} \frac{1}{b^2}
\end{equation}
for s quark.
This replacement follows from the equilibrium of the kinetic energy of 
the quarks in hadrons, i.e.
\begin{equation}
\frac{3}{4m_{\rm u,d} b^2} = \frac{3}{4 m_{\rm s} b_{\rm s}^2} .
\end{equation}
The total wave function of baryon--baryon system is then written as
\begin{equation}
\Psi = {\cal A} [ \varphi_1 \varphi_2 \chi ] ,
\end{equation}
where ${\cal A}$ is the antisymmetrization operator and $\chi$ is the wave
function of the relative motion between two baryon clusters.

For the interaction between quarks, the one-gluon-exchange potential 
(OGEP) of the Fermi-Breit type\cite{Rujula75} is used.
Its form is
\begin{eqnarray}
V_{ij}^{\rm OGEP}(\rbm_{ij}) &=& \lambdabm_i \cdot \lambdabm_j 
\frac{\alpha _{\rm s}}{4}\left[\frac{1}{r_{ij}}-\frac{\pi}{m_i m_j}
\left(1+\frac{2}{3} \sigmabm_i \cdot \sigmabm_j \right)\delta (\rbm_{ij})
\right]
\nonumber \\ & & 
-\lambdabm_i \cdot \lambdabm_j \frac{\alpha _{\rm s}}{4}
\frac{1}{2 m_i m_j} \left( \frac{\pbm_i \cdot \pbm_j}{r_{ij}} + 
\frac{\rbm_{ij} \cdot (\rbm_{ij} \cdot \pbm_i) \pbm_j}{r_{ij}^3} \right) ,
\label{Fermi-Breit}
\end{eqnarray}
where $\rbm_{ij} = \rbm_i - \rbm_j$ is the relative coordinate between quarks, 
and $\pbm_i$ and $\pbm_j$ are the momenta of the quarks.
In the right-hand side of (\ref{Fermi-Breit}), the first term is 
momentum-independent while the second term is the momentum-dependent term.
The part proportional to $\sigmabm_i \cdot \sigmabm_j$ in the first term 
is the color-magnetic interaction (CMI) term, which gives a large attraction 
for the H-dibaryon state.

The flavor SU(3) symmetry breaking (FSB) is implemented in several ways.
It appears in the following parts: 
(i) the wave function; (ii) the quark mass term; (iii) the kinetic term; 
(iv) OGEP.
The FSB in the quark mass term causes only constant shifts.
In order to avoid complications, (i) and (iii) are often treated 
as flavor symmetric, and all the effects from FSB is ascribed to (iv).
The essential results are the same for the baryon spectra and the NY 
scattering phase shifts, independently of the way FSB is taken into account.

Takeuchi and Oka\cite{Takeuchi91,Oka91} studied another type of interquark 
force, which is induced by light-quark--instanton coupling and 
therefore named the instanton induced interaction 
(III).\cite{tHooft76,Shifman80}
III was originally derived by 't Hooft to represent the effect of the 
U$_{\rm A}$(1) symmetry breaking.\cite{tHooft76}
Takeuchi and Oka reduce III to a nonrelativistic form, which consists of 
the two- and three-body forces.\cite{Oka89,Takeuchi91,Oka91,Oka96}
The two-body interaction terms include a color-magnetic term, which 
has the same spin-color structure as that in OGEP.
The three-body interaction has interesting features.
It acts among u-d-s quarks, but in a single baryon it is inactive so 
that only the two-body part of III contributes to the mass splittings 
among baryons.
The three-body part of III acts as a strong short-range repulsive force 
between strange baryons.
If we assume that a part of CMI stems from the two-body part of III, 
the hadron mass spectra are hardly affected except for the 
$\eta$--$\eta'$ splitting.
In the H-dibaryon, however, the short-range repulsion due to the 
three-body part of III may reduce or even cancel the attractive force 
due to CMI.
Unfortunately, ambiguity in determining the strength of the three-body 
force of III prevents us from drawing a definite conclusion.
However, it is interesting to point out that, for the spectrum of dibaryons 
with strangeness, there may be a driving force other than OGEP.

One comment is in order.
Morimatsu and Takizawa\cite{Morimatsu93} estimated the effects of a 
U$_{\rm A}$(1) breaking interaction of the form:\cite{Kobayashi71}
\begin{equation}
{\cal L}_6 = -6 K \left\{ \det ( \bar{\psi}_{\rm R} \psi_{\rm L} ) + 
(\mbox{h.c.}) \right\} ,
\label{UA1}
\end{equation}
where $\psi$ is the quark field and $K$ is a coupling constant determined so 
as to reproduce the mass spectra of the pseudoscalar meson nonet reasonably.
Using the wave functions of the MIT bag model or the nonrelativistic quark 
model, they calculated matrix elements of the U$_{\rm A}$(1) breaking 
interaction for various baryon and two-baryon channels.
They found that for the H-dibaryon the three-body part of the interaction 
(\ref{UA1}) also gives a repulsion, but that its magnitude is much less than 
the result of Takeuchi and Oka.\cite{Takeuchi91,Oka91}
The two-body part gives strong attraction ($-36 \sim -86$ MeV), while the 
three-body part gives somewhat moderate repulsion 
($4 \sim 17$ MeV).\footnote{The instanton contribution on H-dibaryon mass 
is estimated also in the earlier work of Ref.~\citen{Dorokhov92}. 
The contribution from three-body force is very small (about 3 MeV) in that 
calculation.}

Though the OGEP is known to reproduce the short-range part of the NN 
phase shift successfully, the medium- and long-range part needs the 
meson-exchange contribution.
The simplest phenomenological treatment is to include the effective 
meson exchange potential (EMEP) as an interaction potential between 
baryons.\cite{Oka81}
More elaborate studies to include the meson exchange as an interaction 
between quarks have been done although mesons are also composites of 
quark-antiquark.

The earlier study by Oka {\it et al.}\cite{Oka83} focused mainly on the 
short-range part of the baryon--baryon interaction, and the meson-exchange 
was beyond the scope of it.
The first study that applied the meson-exchange potential to the H-dibaryon 
channel was by Straub {\it et al.}\cite{Straub88}
They include the pseudoscalar-meson exchange potential on the quark level,
\begin{eqnarray}
V_{ij}^{\rm psM} &=& -\frac{1}{3} \frac{g_{\rm qqM}^2}{4\pi} 
\frac{\mu^2}{4 M_A M_B} \frac{\Lambda^2}{\Lambda^2-\mu^2} \exp (-\mu^2 b^2 /3)
\nonumber \\
 & & \times 
\mu \left( \frac{\exp (-\mu r_{ij})}{\mu r_{ij}} - \frac{\Lambda^3}{\mu^3} 
\frac{\exp (-\Lambda r_{ij})}{\Lambda r_{ij}} \right) \sigmabm_i \cdot
\sigmabm_j O^{\rm F}_{ij} ,
\end{eqnarray}
where $g_{\rm qqM}$ is the coupling constant, $\mu$ is the meson mass, $M_A$ 
and $M_B$ are the masses of two baryon clusters, $\Lambda$ is the cutoff
parameter and $O^{\rm F}_{ij}$ is the flavor operator, 
and the phenomenological $\sigma$-meson exchange potential on the baryon level,
\begin{eqnarray}
V_{\sigma} (r) = -\frac{g_{\sigma}^2}{4\pi} \frac{1}{2 m_{\sigma}^2
R_{\sigma}^2 r} && \left\{ [1-\exp (-m_{\sigma}r) -\exp (-2m_{\sigma}R_{\sigma}) 
\sinh (m_{\sigma}r)] \theta ( 2R_{\sigma} -r) 
\right.
\nonumber \\
 && \left. \quad 
+ [\cosh (2m_{\sigma}R_{\sigma})
-1 ] \exp (-m_{\sigma}r) \theta (r-2R_{\sigma}) \right\} ,
\end{eqnarray}
where $R_{\sigma}$ is the cutoff radius of the $\sigma$-meson.\cite{Zhang85}
They found that the $\sigma$-meson exchange gives quite important attractive 
contribution.
Only with the one-gluon-exchange potential and the pseudoscalar meson 
exchange, the H-dibaryon does not appear as a bound state, but a resonance 
is found at $E_{\rm cm}=26$ MeV in the $\Lambda\Lambda$-N$\Xi$-$\Sigma\Sigma$ 
three channel phase shift calculation.
When the $\sigma$-meson exchange potential is switched on, the H-dibaryon mass
is pushed down below the $\Lambda\Lambda$ threshold with its binding energy 
$15 \sim 25$ MeV according to the $\sigma$-meson coupling strength 
$g_{\sigma}^2/4\pi$, whose range is determined so as to reproduce the data of 
the $\Lambda$N and $\Sigma$N cross sections. 

Nakamoto {\it et al.} use a more elaborate meson exchange potential, which
incorporates the scalar-meson nonet instead of the flavor-singlet
$\sigma$-meson.\cite{Nakamoto97}
In their $\Lambda\Lambda$-N$\Xi$-$\Sigma\Sigma$ channel coupling calculation 
using one of their models RGM-F, 
an H-dibaryon bound state is found at 19 MeV below the $\Lambda\Lambda$
threshold although this value should be interpreted as a qualitative 
result because in their calculation the mass differences among 
$\Lambda\Lambda$, N$\Xi$ and 
$\Sigma\Sigma$ thresholds are smaller than the experimental values.
Note that the contribution from the flavor-octet scalar-meson exchanges is 
repulsive for the H-dibaryon,\cite{Sakai97} although the net contribution 
from the scalar-meson exchanges is attractive.

Chiral quark models, in which the interaction between constituent quarks is 
mediated by Goldstone bosons, or their hybrid versions describe successfully 
baryon spectra, NN and NY scattering, and properties of the 
deuteron.\cite{Glozman98,Glozman96}\tocite{Zhang97}
Stancu {\it et al.} argued that the H-dibaryon will not bind\cite{Stancu98} 
because the pseudoscalar meson exchange gives rise to a strong repulsion in 
contrast to the one-gluon exchange.
They estimated an adiabatic potential, which is defined as 
$V(R)=\langle H \rangle_R - \langle H \rangle_{\infty}$, where 
$\langle H \rangle_R$ is the expectation value of the hamiltonian with 
respect to the state with the separation distance $R$ between two (0s)$^3$ 
clusters, and $\langle H \rangle_{\infty}=2m_{\Lambda}$.
They obtained $V(0)$=847 MeV and claimed that the H-dibaryon should not 
exist because of this strong repulsion.

Shimizu and Koyama studied more quantitatively the above chiral quark model 
and its hybrid version by an extended RGM, which allows baryons to swell in 
the interaction region.\cite{Shimizu99}
The calculation is made for three cases of the interactions between quarks, 
which reproduce the N$\Delta$ mass difference correctly.
The model (I) uses the OGEP with only long range Yukawa parts of the 
pseudoscalar meson exchange retained.
The model (II) is a hybrid model, in which the pseudoscalar meson exchange 
gives about a third of the N$\Delta$ mass difference and the rest of the 
N$\Delta$ mass splitting is the contribution from the OGEP.
For the meson exchange parts, the pseudoscalar meson exchange (PSME) potential 
$V^{\rm ps}_{ij}(r_{ij})$ and the $\sigma$ meson exchange potential 
$V^{\sigma}_{ij}(r_{ij})$ are taken into account.
They have the following forms:
\begin{eqnarray}
V^{\rm ps}_{ij}(r_{ij}) &=& \frac{1}{3} \frac{g_c^2}{4\pi} 
\frac{m_{\rm ps}^2}{4m_i m_j} \frac{\Lambda^2}{\Lambda^2 -m_{\rm ps}^2} 
\fbm_i \cdot \fbm_j \sigma_i \cdot \sigma_j 
\left\{ \frac{e^{-m_{\rm ps} r_{ij}}}{r_{ij}} - \left( 
\frac{\Lambda}{m_{\rm ps}} \right)^2 \frac{e^{-\Lambda r_{ij}}}{r_{ij}} 
\right\} ,
\nonumber \\
  & & 
\\
V^{\sigma}_{ij}(r_{ij}) &=& \frac{g_c^2}{4\pi} 
\frac{\Lambda^2}{\Lambda^2 -m_{\sigma}^2} 
\left( \frac{e^{-m_{\sigma} r_{ij}}}{r_{ij}} - 
\frac{e^{-\Lambda r_{ij}}}{r_{ij}} \right) ,
\end{eqnarray}
where $\fbm$ is the generators of the flavor SU(3) group.
In model (II), the coupling constant $g_c$ is determined so as to give the 
$\pi$NN coupling constant:
\begin{equation}
\frac{g_c^2}{4\pi} \frac{m_{\pi}^2}{4m_{\rm u,d}^2} = \left( \frac{3}{5} 
\right)^2 \frac{g_{\pi {\rm NN}}^2}{4\pi} \frac{m_{\pi}^2}{4M_{\rm N}^2} .
\end{equation}
The model (III) has no OGEP part and the N$\Delta$ mass difference is wholly 
given by the PSME.
The three coupling constants, $g_c(\pi)$, $g_c(K)$ and $g_c(\eta)$, for 
pseudoscalar meson exchanges are determined so as to give the N$\Delta$ mass 
difference and the thresholds N$\Xi$ and $\Sigma\Sigma$ correctly, whereas 
a common coupling constant is used in models (I) and (II).
The results of the $\Lambda\Lambda$-N$\Xi$-$\Sigma\Sigma$ coupled channel 
calculation are as follows.
When the same $\sigma$-meson exchange potential is used for three models, the 
binding energies for model (I) and (II) are 74.5 MeV and 24.4 MeV, 
respectively, but there is no bound state for model (III).
However, the model (II) and (III) are found to give less attractive NN 
potential.
If the strengths of the $\sigma$-meson exchange potential are increased so as 
to reproduce the phase shifts of the NN $^1$S$_0$ state, the binding energy 
for the model (II) becomes 56.8 MeV and even the model (III) gives a bound 
state with its binding energy 15.6 MeV.
After all, the binding energies of the H-dibaryon in chiral (or hybrid chiral) 
quark models are much less than those in OGE models due to the repulsion 
mainly from the pion exchange, but room for the bound H-dibaryon still remains 
on account of the medium range attraction from the $\sigma$-meson exchange.
The calculation also shows that the wave function of the H-dibaryon is more 
spread out than the simple (0s)$^6$ configuration due to the medium range 
attraction.
The $\Sigma\Sigma$ component is drawn inside because this channel gains a 
lot of attractions at short distances due to the CMI, whereas the 
$\Lambda\Lambda$ and N$\Xi$ components are slightly apart from each other 
to reduce the relative kinetic energy.
A recent study in a hybrid chiral quark model also gives similar 
results.\cite{Shen99}

Quark confinement is realized in the QCMs by a potential among quarks.
Mostly used one is the linear or quadratic type of potential between 
two quarks:
\begin{subequations}
\begin{equation}
 -a_1 | \rbm_1 - \rbm_2 | \lambdabm_1 \cdot \lambdabm_2 \quad \mbox{(linear)}
\end{equation}
or
\begin{equation}
 -a_2 ( \rbm_1 - \rbm_2 )^2 \lambdabm_1 \cdot \lambdabm_2 \quad 
\mbox{(quadratic)}.
\label{conf-quad}
\end{equation}
\end{subequations}
These types of confinement potential hardly affect the physical observables 
of the baryon--baryon system such as the binding energy (of two baryon 
system) or the phase shift.
This is because in the flavor symmetric limit $\langle \sum_{i < j} \lambdabm_i
\cdot \lambdabm_j \rangle$ is dependent only on the number of quarks so that 
the confinement potential term does not give any net force between baryon 
clusters.
Even when the flavor symmetry breaking is introduced in the 
wave function, only negligible contribution arises.

Another quark confinement mechanism called flip-flop model is 
proposed\cite{Lenz86} and
studied in $S=-2$ channel.\cite{Koike90}
The flip-flop model avoids the long-range ``color van der Waals
force"\cite{Willey78} which appears through a virtual excitation of the 
hidden color state (the color octet dipole state).
Such a force cannot be represented by a two-body potential and is necessarily 
a many-body force.
Let us illustrate a two-baryon system to explain the flip-flop model.
First, six quarks are divided into nearest-neighbor three quark systems.
Then if each cluster is in the color-singlet state, both of three quark 
systems are confined separately, i.e. the confining force is inactive 
between the quarks in the different clusters.
Otherwise the confining force acts among all pairs of six quarks.

Including the pseudoscalar and $\sigma$-meson exchange, Koike {\it et al.} 
found the 
H-dibaryon binding energy 64.4$\sim$124.5 MeV varying with the coupling 
constant for the $\sigma$-meson.\cite{Koike90}
This binding energy is rather large compared with other QCM calculations 
with the confinement through the two-body potential.
There is a bound state for another channel with $S=-2$, i.e. $J=1$, $T=0$
N$\Xi$ channel with its binding energy 6.3 MeV, which can decay only via 
weak interaction.

To summarize the QCM for the H-dibaryon, we can extract some common results 
as follows.
Here we consider the calculations using (OGEP)+(meson-exchange
potential)+(two-body confinement potential) and including the FSB as a
``standard calculation" in QCM.

(i) The CMI in the OGEP gives large attractive force but it is not sufficient
for the H-dibaryon to bind.
The scalar meson ($\sigma$-meson) exchange is indispensable.
The coupled channel $\Lambda\Lambda$-N$\Xi$-$\Sigma\Sigma$ calculation without
$\sigma$-meson exchange potential shows a resonance in the $\Lambda\Lambda$
phase shift,\cite{Oka83,Straub88} which disappears in the calculation with 
$\sigma$-meson exchange.\cite{Straub88}


(ii) The FSB reduces the attractive force due to the
CMI.\cite{Oka83,Silvestre87,Nakamoto97}
In the flavor SU(3) symmetric limit, the H-dibaryon bound states appear even
without the $\sigma$-meson exchange.
These bound states vanish when the effect of the FSB is taken into account in 
the calculation without $\sigma$-meson exchange.\cite{Oka83,Silvestre87}

(iii) The $\Lambda\Lambda$-N$\Xi$-$\Sigma\Sigma$ coupled, flavor-singlet
structure is essential for the H-dibaryon to bind.
The $\Lambda\Lambda$ single-channel calculation shows that the interaction
between two $\Lambda$'s has repulsive core.\cite{Oka83,Straub88,Nakamoto97}
On the other hand, 
the N$\Xi$ and $\Sigma\Sigma$ single-channel calculations show the attractive 
interactions in these channels.
This attractive nature leads to 
a resonance structure in the calculation without $\sigma$-meson
exchange\cite{Oka83,Straub88} or a bound state\cite{Straub88} in
the full three-channel calculation.

(iv) Although the relative magnitudes of the $\Lambda\Lambda$, N$\Xi$ and 
$\Sigma\Sigma$ components of the bound H-dibaryon wave function roughly agree 
with those of the flavor-singlet state of 
eq.(\ref{HSU3wf}),\cite{Oka83,Straub88,Koike90} the wave function changes as 
a reflection of the natures of each components.\cite{Shimizu99}

(v) The quark-quark interactions other than OGEP, i.e. III and Goldstone 
boson exchange, may reduce the attractive force in the H-dibaryon channel 
substantially.\cite{Takeuchi91,Oka91,Stancu98,Shimizu99,Shen99}


\section{Nucleon--H-dibaryon interaction}
\label{sec:NH}

In this section, we will review a study on the interaction between a 
nucleon and an H-dibaryon in Ref.~\citen{Sakai92} and further comment on the 
possible implication of the H-dibaryon to the world of the nucleus with $S=-2$.
As stated in the introduction, the H-dibaryon is not only an interesting 
object in itself but also important in $S=-2$ sector nuclear 
physics.
In fact, though a few events of double hypernuclei were 
reported\cite{Danysz63,Prowse66,Aoki91b} and several candidate events have 
successively been reported recently,\cite{Tanaka95,Nakazawa95} structures of 
these double hypernuclei have not been fully understood yet.
It is possible that there is a double hypernucleus which have the character of 
an H-nucleus 
rather than $\Lambda\Lambda$ nucleus, if the H-dibaryon is strongly bound in 
the nucleus.

In \S\ref{subsec:NH-1}, the framework of the QCM commonly used in the
calculation of the NH and HH (\S\ref{sec:HH}) interaction is explained.
The results for the NH interaction\cite{Sakai92} are recapitulated in 
\S\ref{subsec:NH-2}.
In \S\ref{subsec:NH-3}, a study on the behavior of the H-dibaryon in nuclear
matter\cite{Sakai95} is reviewed.
The relation with the $\Lambda\Lambda$ in nuclear matter is also discussed.

\subsection{The framework of the quark cluster model}
\label{subsec:NH-1}

In this work and the work on the interaction between two H-dibaryons in the
next section, the following framework of the quark cluster model is used.

(i) The orbital parts of the internal wave functions of the nucleon and 
the H-dibaryon have (0s)$^3$ and (0s)$^6$ configurations, respectively, 
and the flavor symmetry breaking (FSB) is not introduced 
in the wave function, i.e. FSB is reflected only in the hamiltonian.
The internal wave functions of a nucleon $\phi _{\rm N}(\xibm_{\rm N})$ and an 
H-dibaryon $\phi _{\rm H}(\xibm _{\rm H})$ are then written as
\begin{eqnarray}
\phi _{\rm N}(\xibm_{\rm N}) &=& \varphi _{\rm N}(\xibm_{\rm N}){\cal C}
([111]_{\rm C}){\cal S}([3]_{\rm SF})  ,
\nonumber  \\
\phi _{\rm H}(\xibm _{\rm H}) &=& \varphi _{\rm H}(\xibm _{\rm H})
{\cal C}([222]_{\rm C}){\cal S} ([222]_{\rm SF}) ,
\end{eqnarray}      
where $\xibm _{\rm N}$ and $\xibm _{\rm H}$ are internal coordinates.
Here, $\varphi _{\rm N}$ and $\varphi _{\rm H}$ are the orbital parts, 
${\cal C}$ is the color part and ${\cal S}$ is the spin-flavor part of the 
internal wave functions.

For the relative motion, only S-waves are considered.

(ii) The hamiltonian is
\begin{equation}
H=K+V ,
\label{eq:hamil}
\end{equation}
where $K$ is the sum of the kinetic energy of each quark with the 
center-of-mass kinetic energy of the totalsystem subtracted, 
\begin{equation}
K=\sum_{i=1}^{N_q} \frac{\pbm _i^2}{2m_i}-\frac{(\sum \pbm_i)^2}{2 \sum m_i} ,
\end{equation}
where $N_q$ is the total number of quarks, 
and $V$ is the potential term which consists of the residual interaction and
the confinement term $V^{\rm conf}$.
For the residual interaction, the momentum-independent part of the 
one-gluon-exchange potential of the Fermi-Breit type $V_{ij}^{\rm OGEP}$
(\ref{Fermi-Breit}) supplemented by meson exchange potential is used.
The relevant part of $V_{ij}^{\rm OGEP}$ is
\begin{equation}
V_{ij}^{\rm OGEP}=\lambdabm_i \cdot \lambdabm_j 
\frac{\alpha _{\rm s}}{4}\left[\frac{1}{r_{ij}}-\frac{\pi}{m_i m_j}
\left(1+\frac{2}{3} \sigmabm_i \cdot \sigmabm_j \right)\delta (\rbm_{ij})
\right] .
\end{equation}
Here, $\alpha _{\rm s}$ is the quark-gluon coupling constant.
The color magnetic interaction term plays important role in producing the
H-dibaryon bound state and short-range repulsive forces between NN, NH or HH.
So FSB is introduced by the following parametrization:
\begin{equation}
\frac{\pi}{m_i m_j}\rightarrow \xi_{ij}\frac{\pi}{{m_u}^2}
\label{eq:FSB}
\end{equation}
so that the thresholds, $\Lambda \Lambda $, $N\Xi$ and $\Sigma \Sigma$ are
correctly given.\cite{Oka83}
Here, the parameter $\xi_{ij}=1$ when both {\it i} and {\it j} are u- or 
d-quarks,
$\xi_{ij}=\xi_1$ when either {\it i} or {\it j} is an s-quark, and 
$\xi_{ij}=\xi_2$ when both {\it i} and {\it j} are s-quarks.
In the naive interpretation of (\ref{eq:FSB}), $\xi_2=\xi_1^2$.
However, as a consequence that all FSB effects are burdened to this term, the
actual values are taken to be $\xi_1=0.6$ and $\xi_2=0.1$.\cite{Oka83}
For the two-body confinement potential, quadratic type (\ref{conf-quad}) is
used, i.e. 
\begin{equation}
V^{\rm conf} = -a_2 \sum_{i<j} r_{ij}^2 \lambdabm_i \cdot \lambdabm_j .
\end{equation}

(iii) The meson exchange contributions to NH and HH interactions are introduced
as follows.
An effective meson exchange potential (EMEP) is introduced into the RGM
hamiltonian kernel,\cite{Oka81} 
\begin{equation}
{V}^{\rm EMEP}_{\rm RGM} (\Rbm , \Rbm ') = \int N^{1/2}_{\rm RGM} (\Rbm ' , 
\Rbm ''){\cal V}(R'')N^{1/2}_{\rm RGM}(\Rbm '',\Rbm ){\rm d}\Rbm '' ,
\label{eq:emep}
\end{equation}
where the phenomenological potential ${\cal V}(R)$ is taken to be
gaussian\cite{Sakai95,Sakai97}
\begin{equation}
{\cal V}(R)=V_0 \exp (-R^2/\alpha ^2).
\label{eq:phen}
\end{equation}
This is a simple extension of the method developed in the pioneering work of
the NN interaction in the quark cluster model.\cite{Oka81}
First, the range of the strength of EMEP for the H-dibaryon, $V_0^{\rm H}$, 
is determined to give the mass of the
H-dibaryon between $\Lambda\Lambda$ threshold and the lower limit determined 
from the KEK experiment,\cite{Aoki91b} i.e. 
\begin{equation}
0<2M_\Lambda-M_{\rm H} < 27.6~\mbox{MeV} .
\label{eq:masscon}
\end{equation}
Then, the gaussian size parameters, $\alpha_{\rm NH}$ and $\alpha_{\rm HH}$, 
and the strength of EMEP, $V_0^{\rm NH}$ and $V_0^{\rm HH}$, for the NH and 
HH interactions, respectively, are obtained by the 
direct convolution\cite{Sakai92,Sakai97} of the EMEP for the H-dibaryon as
\begin{subeqnarray}
& & \alpha_{\rm NH}^2 = \alpha_{\rm H}^2 + \frac{1}{6} b^2 , \qquad 
V_0^{\rm NH} = 2 V_0^{\rm H} \left( \frac{\alpha_{\rm H}}{\alpha_{\rm NH}} 
\right)^3 ,
\\
& & \alpha_{\rm HH}^2 = \alpha_{\rm H}^2 + \frac{1}{3} b^2 , \qquad 
V_0^{\rm HH} = 4 V_0^{\rm H} \left( \frac{\alpha_{\rm H}}{\alpha_{\rm HH}} 
\right)^3 .
\end{subeqnarray}

The parameters used in the calculations are listed in
Table~\ref{tab:parameters}.
\begin{table}
\caption{The parameters used in the calculation of the NH and HH interactions.}
\label{tab:parameters}
\begin{center}
\begin{tabular}{cccccccc} \hline \hline
$m$ & $b$ & $\alpha_s$ & $a_2$ & $\xi_1$ & $\xi_2$ & $\alpha_{\rm NH}$ & 
$\alpha_{\rm HH}$ \\
(MeV) & (fm) &   & $\mbox{(MeV/fm}^2)$ &   &   & (fm) & (fm) \\ \hline
300 & 0.6 & 1.39 & 33.0 & 0.6 & 0.1 & 0.97 & 1.00 \\ \hline \hline
$2M_{\Lambda}-M_{\rm H}$ & $V_0^{\rm NH}$ & $V_0^{\rm HH}$ &  &  &  &  &  \\
                         & (MeV)         & (MeV)           &  &  &  &  &  \\
\hline
0                        & $-601$        & $-1096$         &  &  &  &  &  \\
8.5                      & $-623$        & $-1136$         &  &  &  &  &  \\
27.6                     & $-673$        & $-1227$         &  &  &  &  &  \\
43.4                     & $-714$        & $-1302$         &  &  &  &  &  \\
\hline  
\end{tabular}
\end{center}
\end{table}

\subsection{The property of the NH interaction}
\label{subsec:NH-2}

\begin{figure}
\parbox[t]{\halftext}{
\epsfxsize=72mm
\centerline{\epsfbox{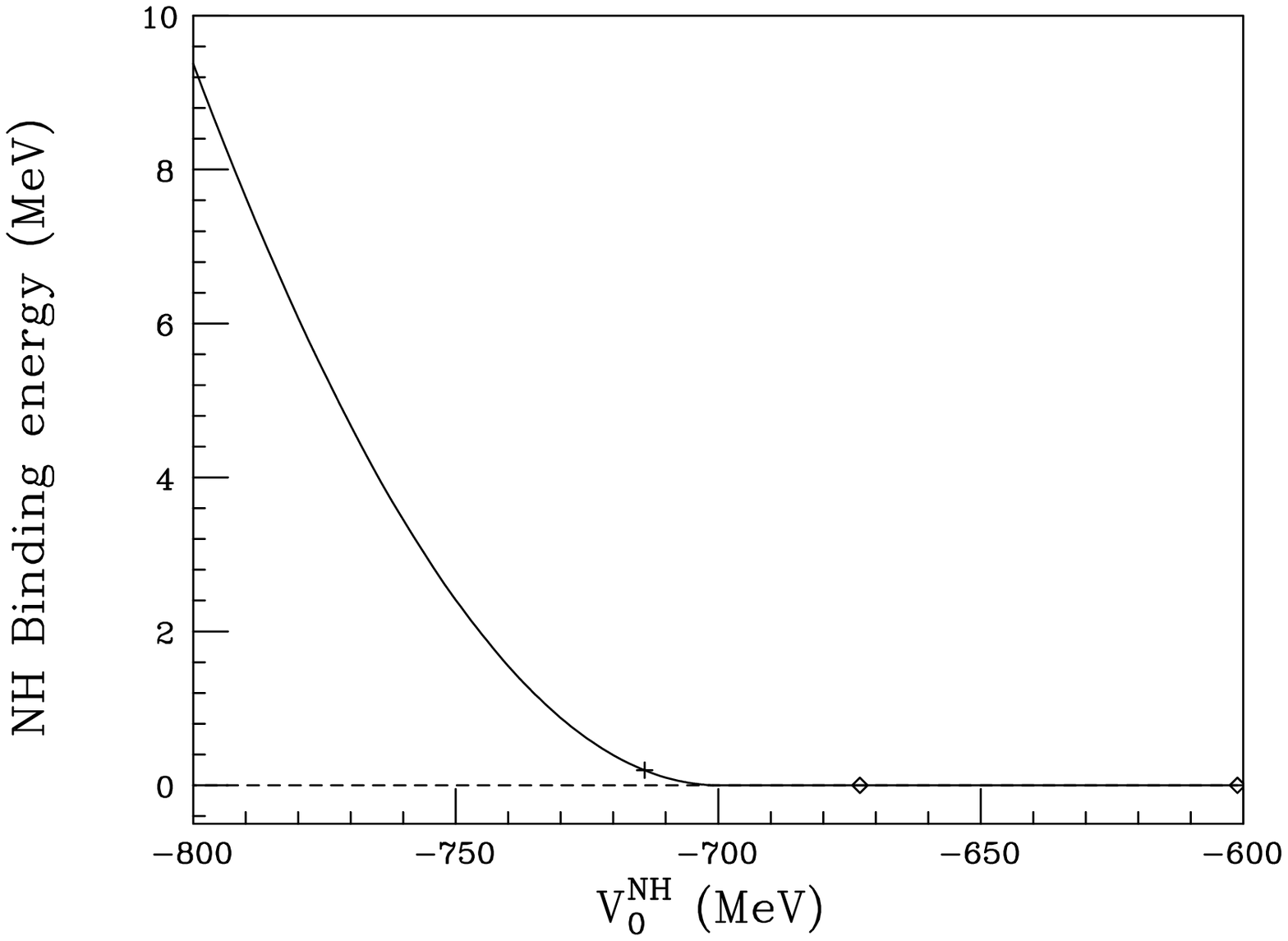}}
\caption{The binding energy of the NH system as a function of the strength of
EMEP $V_0^{\rm NH}$. (Taken from Ref.~{\protect\citen{Sakai92}}.)}
\label{fig:NH-BE}
}
\hspace{6mm}
\parbox[t]{\halftext}{
\epsfxsize=72mm
\centerline{\epsfbox{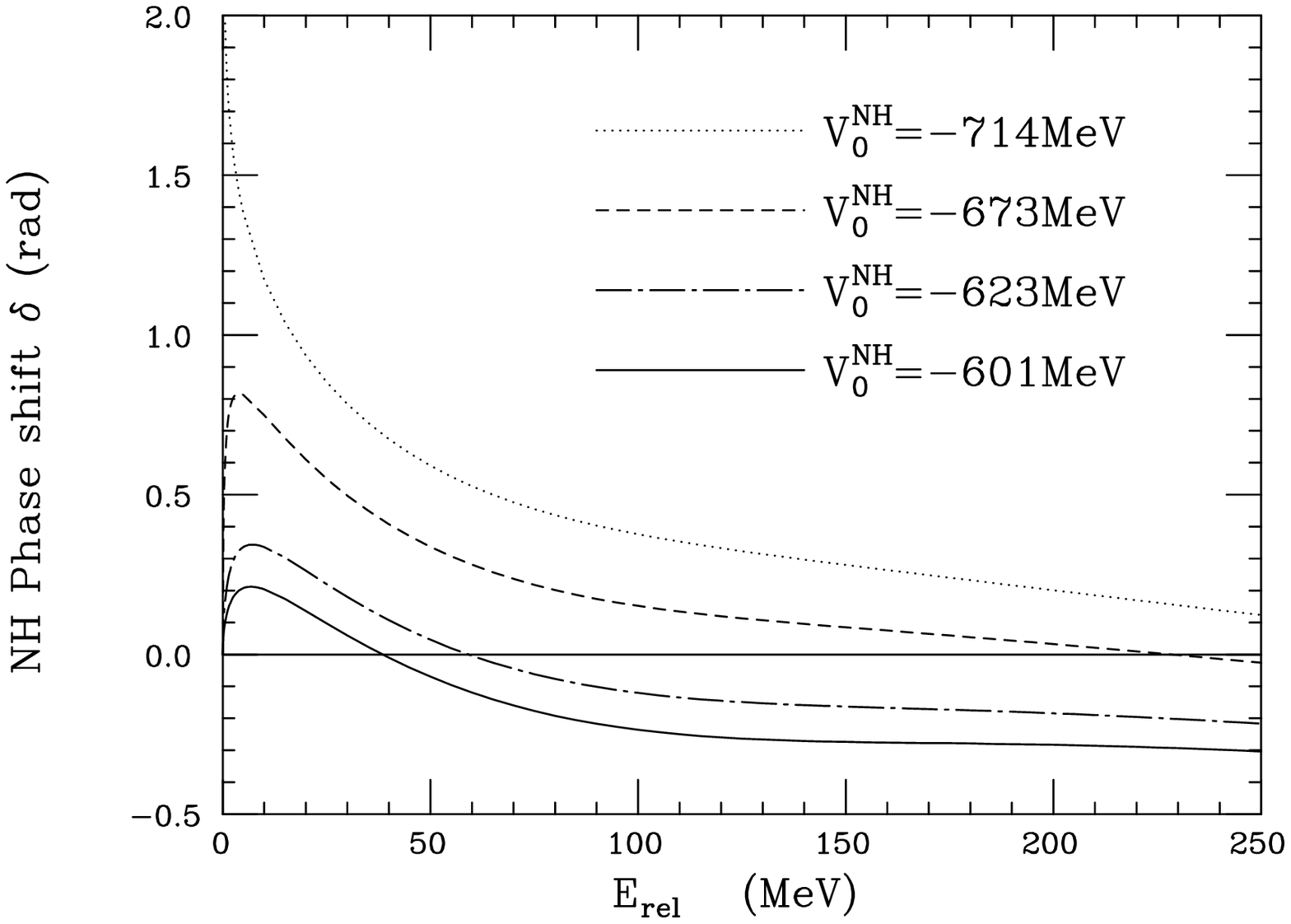}}
\caption{The NH scattering phase shift $\delta$ as a function of the 
relative energy $E_{\rm rel}$ for several $V_0^{\rm NH}$. 
(Taken from Ref.~{\protect\citen{Sakai92}}.)}
\label{fig:NH-ps}
}
\end{figure}

In Fig.~\ref{fig:NH-BE} the binding energy of the NH system as a function of 
the strength parameter of EMEP, $V_0^{\rm NH}$, is shown.
The NH scattering phase shift $\delta$ is shown in Fig.~\ref{fig:NH-ps} as 
a function of the relative energy $E_{\rm rel}$ for several $V_0^{\rm NH}$.
The essential results of this calculation are as follows.

(i) The bound state exists only for the rather strong EMEP.
For example, NH system is loosely bound with its binding energy 0.2 MeV for
$V_0^{\rm NH}=-714$ MeV.

(ii) The fundamental character of the NH interaction is that it has a 
repulsive core in the short-range region and an attractive force in the 
medium-range region.
This nature of the interaction is clearly seen in the behavior of the phase
shift of Fig.\ref{fig:NH-ps}.
The origin of the short-range repulsion is the Pauli principle for quarks and 
CMI.
The normalization kernel, which represents the overlap of the wave function in 
a sense, is $\frac{5}{8}$ at zero separation of NH clusters, while one at 
infinity.
Therefore the Pauli blocking is $\frac{3}{8}$ times as effective as the
complete blocking.
The role of CMI can be qualitatively explained as follows.
In the flavor SU(3) symmetric limit, the CMI term is proportional to $\Theta$ 
defined in (\ref{CMI-op}).
For a nucleon and an H-dibaryon, $\Theta_{\rm N}=-8$ and $\Theta_{\rm H}=-24$, 
respectively, while for an NH system of zero separation $\Theta_{\rm NH}=4$.
The difference $\Theta_{\rm NH} - (\Theta_{\rm N}+\Theta_{\rm H}) = 36$ causes 
a strong repulsive force.
The main contribution to the medium-range attractive force is from the
$\sigma$-meson exchange.

\subsection{H-dibaryon in nuclear matter}
\label{subsec:NH-3}

In Ref.~\citen{Sakai95}, the property of the H-dibaryon in nuclear matter 
is investigated.
First, the single-particle potential of the H-dibaryon in nuclear matter is 
calculated employing the Brueckner theory.
The non-local potential between N and H derived from the QCM
(\S\ref{subsec:NH-2}) is Fourier-transformed and then used to solve the 
Bethe--Goldstone equation.
In this work, N and H are treated as elementary particles.
Single-particle potential in uniform matter is well described by an effective 
mass and a depth of the potential.
The effective mass of the H-dibaryon, $M^{\ast}_{\rm H}$, and the potential 
well depth, $D_{\rm H}$, are dependent on the EMEP parameter $V_0^{\rm NH}$.
The obtained values are listed in Table~\ref{tab:HinNM}.
\begin{table}[htb]
\caption{The ratio of the H-dibaryon effective mass in nuclear matter, 
$M_{\rm H}^{\ast}$, to the H-dibaryon mass in free space, $M_{\rm H}$, and 
the well depth, $D_{\rm H}$, in nuclear matter.
$V_0^{\rm NH}$ is the strength of the effective meson exchange potential 
between a nucleon and an H-dibaryon.
Those under the effect of the coupling with $\Lambda\Lambda$, 
$M_{\rm H}^{\ast {\rm (c)}}$ and $D_{\rm H}^{\rm (c)}$, are also shown for 
the case $\Delta M_{\rm H}=0$. }
\label{tab:HinNM}
\begin{center}
\begin{tabular}{rrrrr}
\hline \hline
$V_0^{\rm NH}$ (MeV) & $M_{\rm H}^{\ast} / M_{\rm H}$ & $D_{\rm H}$ (MeV) 
& $M_{\rm H}^{\ast {\rm (c)}} / M_{\rm H}$ & $D_{\rm H}^{\rm (c)}$ (MeV) \\
\hline
$-601$ &  0.853 &  14.5 & 0.848 &  7.2 \\
$-623$ &  0.820 &  28.7 & 0.814 & 22.3 \\
$-673$ &  0.769 &  63.6 & 0.769 & 60.0 \\
$-714$ &  0.755 &  95.3 & 0.750 & 90.1 \\
\hline
\end{tabular}
\end{center}
\end{table}

\begin{figure}
\parbox[t]{\halftext}{
\epsfxsize=76mm
\centerline{\epsfbox{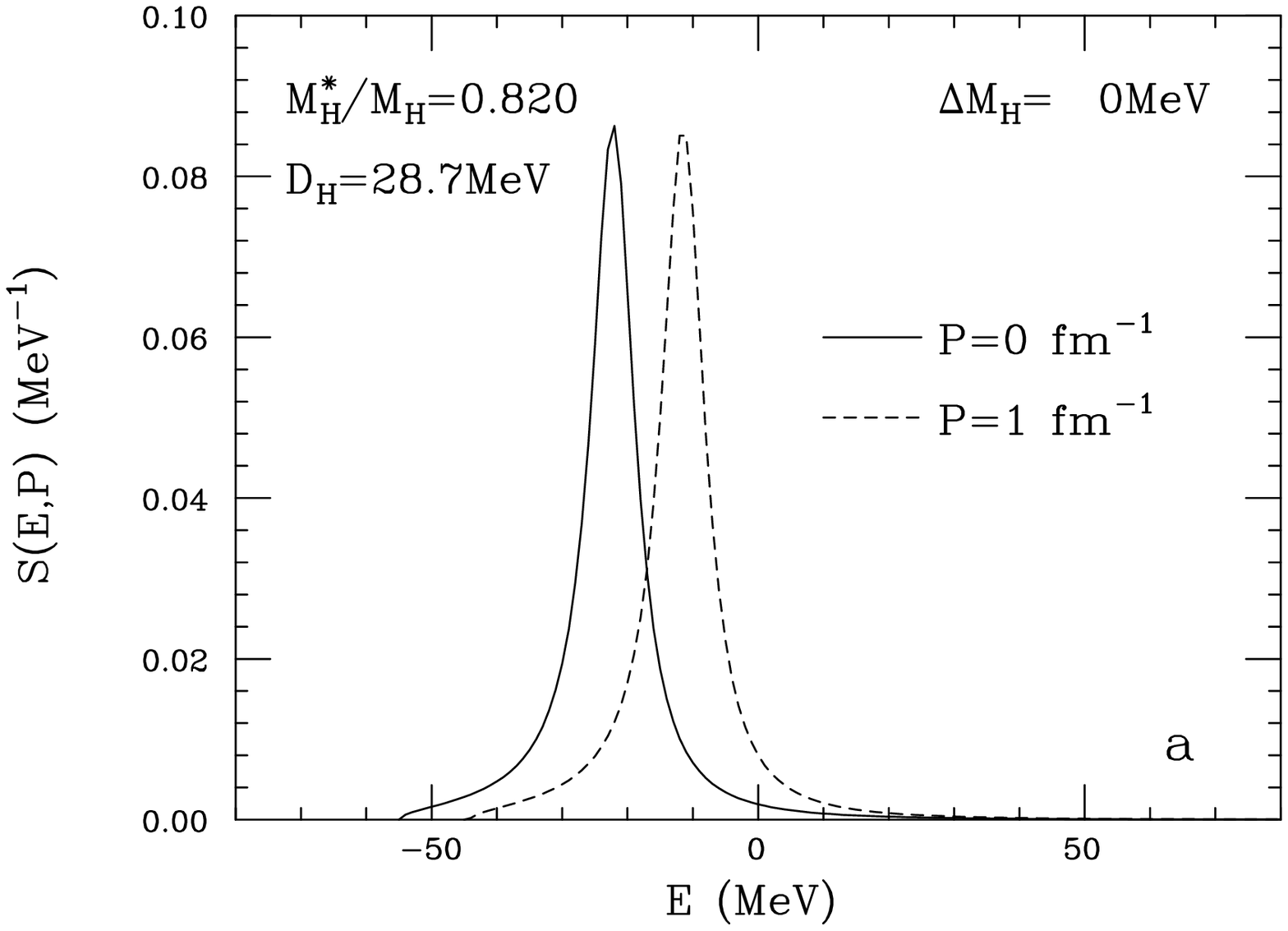}}
}
\hspace{6mm}
\parbox[t]{\halftext}{
\epsfxsize=76mm
\centerline{\epsfbox{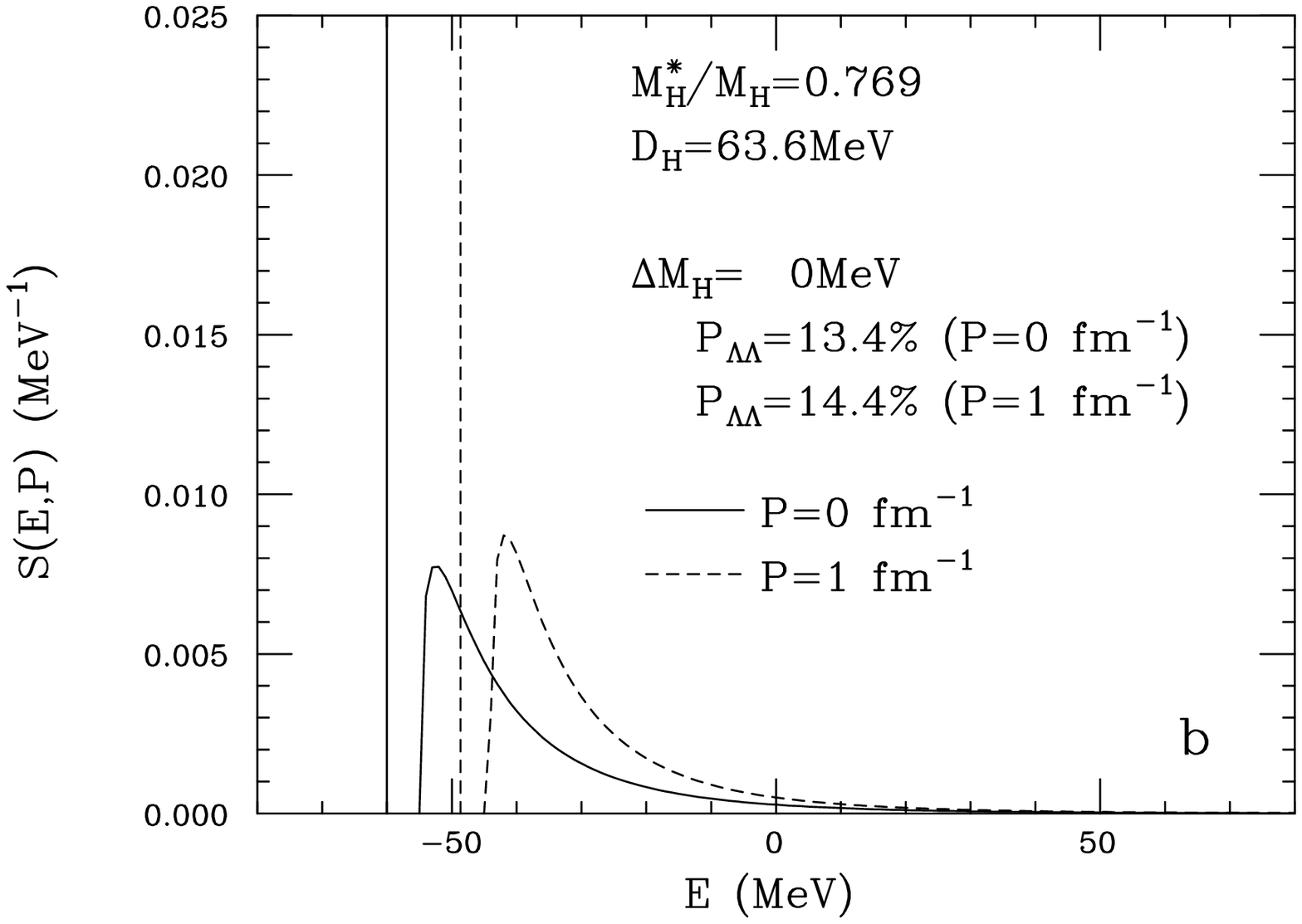}}
}
\caption{
(a) The spectral function, $S(E,P)$, at $P=0$ and 1 fm$^{-1}$ is shown for
$M_{\rm H}^{\ast}/M_{\rm H} = 0.820$ and $D_{\rm H} = 28.7$ MeV.
$\Delta M_{\rm H} = 0$.
The solid (broken) line corresponds to
$P=0$ fm$^{-1}$ ($P=1$ fm$^{-1}$).
(b) Same as (a) but for
$M_{\rm H}^{\ast}/M_{\rm H} = 0.769$ and $D_{\rm H} = 63.6$ MeV.
The amounts of the $\Lambda\Lambda$ admixture in bound states,
$P_{\Lambda\Lambda}$, are also given.
(Data in these figures are taken from Ref.~{\protect\citen{Sakai95}}.)}
\label{fig:spectral}
\end{figure}

In the actual situation, the coupling with the $\Lambda\Lambda$ channel is
important for the H-dibaryon near the $\Lambda\Lambda$ threshold.
This effect is taken into account by a model with a simple coupling vertex
function between H-dibaryon and $\Lambda\Lambda$, 
\begin{equation}
\Gamma ( k ) = g e^{-b^2 k^2} .
\end{equation}
The physical H-dibaryon mass $M_{\rm H}$
($\Delta M_{\rm H}=M_{\rm H}-2 M_{\Lambda}$) is determined by the 
position of the pole of the propagator:
\begin{equation}
\Delta M_{\rm H} - \Delta M_{\rm H}^{(0)}
- {\rm Re} [ g^2 \Sigma ( \Delta M_{\rm H} ) ] = 0 ,
\label{propagator-pole}
\end{equation}
where $\Delta M_{\rm H}^{(0)}$ is the bare H-dibaryon mass measured from 
the $\Lambda \Lambda$ threshold and $\Sigma$ is the self-energy defined 
by
\begin{equation}
g^2 \Sigma ( E ) = \int \frac{{\rm d} \qbm}{(2 \pi )^3}
\frac{\Gamma^2(q)}{E - q^2 / M_{\Lambda} + i \epsilon } .
\end{equation}
Eq.(\ref{propagator-pole}) determines $\Delta M_{\rm H}$ provided 
$\Delta M_{\rm H}^{(0)}$ and $\Gamma ( q )$ are given.
Here, inversely, eq.(\ref{propagator-pole}) is used to determine the bare 
mass $M_{\rm H}^{(0)}$ from a given $\Delta M_{\rm H}$.
$\Gamma (k)$ is determined so as to reproduce the low-energy 
behavior of $\Lambda \Lambda$ scattering given by a QCM in 
Ref.~\citen{Oka91}.
Then using $M_{\rm H}^{(0)}$ and $\Gamma (k)$, the properties of the 
H-dibaryon in nuclear matter affected by the coupling with the 
$\Lambda\Lambda$ channel are studied through the analysis of the 
H-dibaryon propagator in nuclear matter, $G_{\rm H} (E, P)$:
\begin{equation}
G_{\rm H}^{-1} (E, P) =
E - \frac{P^2}{2 M_{\rm H}^{\ast} }
- ( \Delta M_{\rm H}^{(0)} - D_{\rm H} )
- \int \frac{{\rm d} \qbm}{(2 \pi )^3}
\frac{\Gamma ( q ) ^2}{ E - \frac{P^2}{4 M_{\Lambda}^{\ast} }
- \frac{q^2}{M_{\Lambda}^{\ast} } + 2 D_{\Lambda} + i \epsilon } ,
\end{equation}                            
where $P$ is the momentum of the H-dibaryon.
The potential well depth of $\Lambda$ in nuclear matter,
$D_{\Lambda}$, used here is 27.5 MeV and 
$M_{\Lambda}^{\ast}/M_{\Lambda}=0.8$.\cite{Chrien89}
Here, $M_{\rm H}^{\ast}/M_{\rm H}$ and $D_{\rm H}$ are regarded as the 
quantities for the ``bare'' H-dibaryon, which are not subject to the coupling.
The total energy (= kinetic energy + potential energy) of
the H-dibaryon in nuclear matter, $E_{\rm H} ( P )$, obtained from the 
condition
\begin{equation}
{\rm Re} [ G_{\rm H}^{-1} (E_{\rm H}, P) ] = 0  ,
\end{equation}
is parametrized in the effective mass approximation:
\begin{equation}
E_{\rm H} (P) = \frac{P^2}{2 M_{\rm H}^{\ast {\rm (c)}} }
- D_{\rm H}^{{\rm (c)}} .
\end{equation}
$M_{\rm H}^{\ast {\rm (c)}}/M_{\rm H}$ and $D_{\rm H}^{{\rm (c)}}$ 
for $\Delta M_{\rm H}=0$ are also shown in Table~\ref{tab:HinNM}.
The value of $D_{\rm H}^{{\rm (c)}}$ is to be compared with $2 D_{\Lambda}$.
If the energy of the H-dibaryon, 
$\frac{P^2}{2 M_{\rm H}^{\ast {\rm (c)}}}-D_{\rm H}^{{\rm (c)}}$, is 
less than the $\Lambda\Lambda$ threshold energy, 
$2 \left[ \frac{(P/2)^2}{2 M_{\Lambda}^{\ast}}-D_{\Lambda} \right]$, in 
nuclear matter, the H-dibaryon becomes the ground state of an $S=-2$ 
two-baryon system in nuclear matter.
From the value of $D_{\rm H}^{{\rm (c)}}$ listed in Table~\ref{tab:HinNM}, 
the H-dibaryon does not appear as the bound state in nuclear matter and 
the ground state in nuclear matter is $\Lambda\Lambda$ for most of the
plausible values of $V_0^{\rm NH}$.
Only for relatively strong EMEP, the H-dibaryon appears as a bound state.
However, even when the ground state in nuclear matter is 
$\Lambda\Lambda$, the continuum above the threshold is a mixed state of 
$\Lambda\Lambda$ and H-dibaryon, and there is a region where the H-dibaryon 
component becomes strong.
To see this, the spectral function 
\begin{equation}
S(E,P) = -\frac{1}{\pi} {\rm Im} [ G_{\rm H} (E,P) ]
\end{equation}
is shown in Fig.\ref{fig:spectral}.
Fig.\ref{fig:spectral}a corresponds to the case where the energies of the 
H-dibaryon exceed the $\Lambda\Lambda$ thresholds in nuclear matter, and 
shows that narrow ranges around peaks contain appreciable amounts of the
H-dibaryon component.
For comparison, shown in Fig.\ref{fig:spectral}b is the case where the 
H-dibaryon appears as the bound state, which is seen as the 
$\delta$-function peak in the spectral function.

Though this work has been done for nuclear matter, it may give an 
implication to finite double hypernuclei.
The situation for a finite nucleus, corresponding to the case where there 
is no H-bound state in nuclear matter, will be that the
low-lying discrete states have the character of $\Lambda\Lambda$ bound
states, but that some of excited-states may have strong admixture of the 
H-nuclear states.
In the opposite case where H is bound below the $\Lambda\Lambda$-threshold in
nuclear matter, it is expected that the ground states in finite nuclei with 
strangeness $-2$ will have the character of the H-nuclear states and the 
amounts of the mixing with $\Lambda\Lambda$ states are roughly given by the 
nuclear matter calculation.

\section{Interaction between H-dibaryons}
\label{sec:HH}

In this section, we will review an investigation on the 
interaction between two H-dibaryons in Ref.~\citen{Sakai97}.
As the framework of the QCM is the same as in the work reviewed in the previous
section, we do not repeat here.

Tamagaki suggested the possibility that H-matter appears at densities 
several times higher than normal nuclear density.\cite{Tamagaki91}
That work is based on an assumption that the CMI plays a key role in 
determining the properties of the H--H interaction.
In this pioneering study, a simple H--H interaction model consisting 
of a hard-core potential plus a square well attractive potential 
outside the core was used, because no microscopic calculation of the 
H--H interaction was available at that time.
In the work reviewed here,\cite{Sakai97} more quantitative information on 
the H--H interaction is extracted from a microscopic calculation by 
employing the quark cluster model.

\begin{figure}
\parbox[t]{\halftext}{
\epsfxsize=72mm
\centerline{\epsfbox{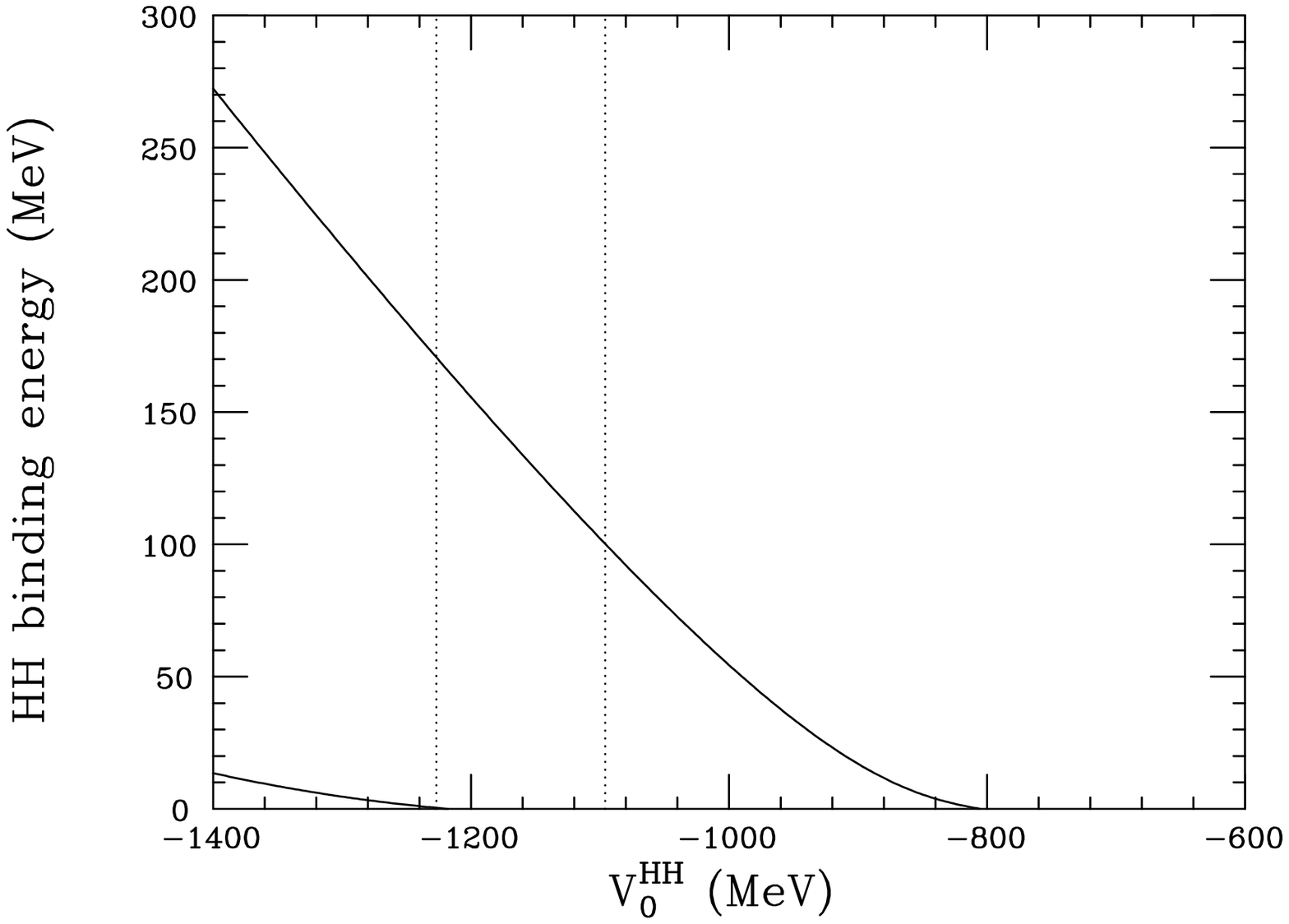}}
\caption{The binding energy of the HH system as a function of the strength of
EMEP $V_0^{\rm HH}$. (Taken from Ref.~{\protect\citen{Sakai97}}.)}
\label{fig:HH-BE}
}
\hspace{6mm}
\parbox[t]{\halftext}{
\epsfxsize=72mm
\centerline{\epsfbox{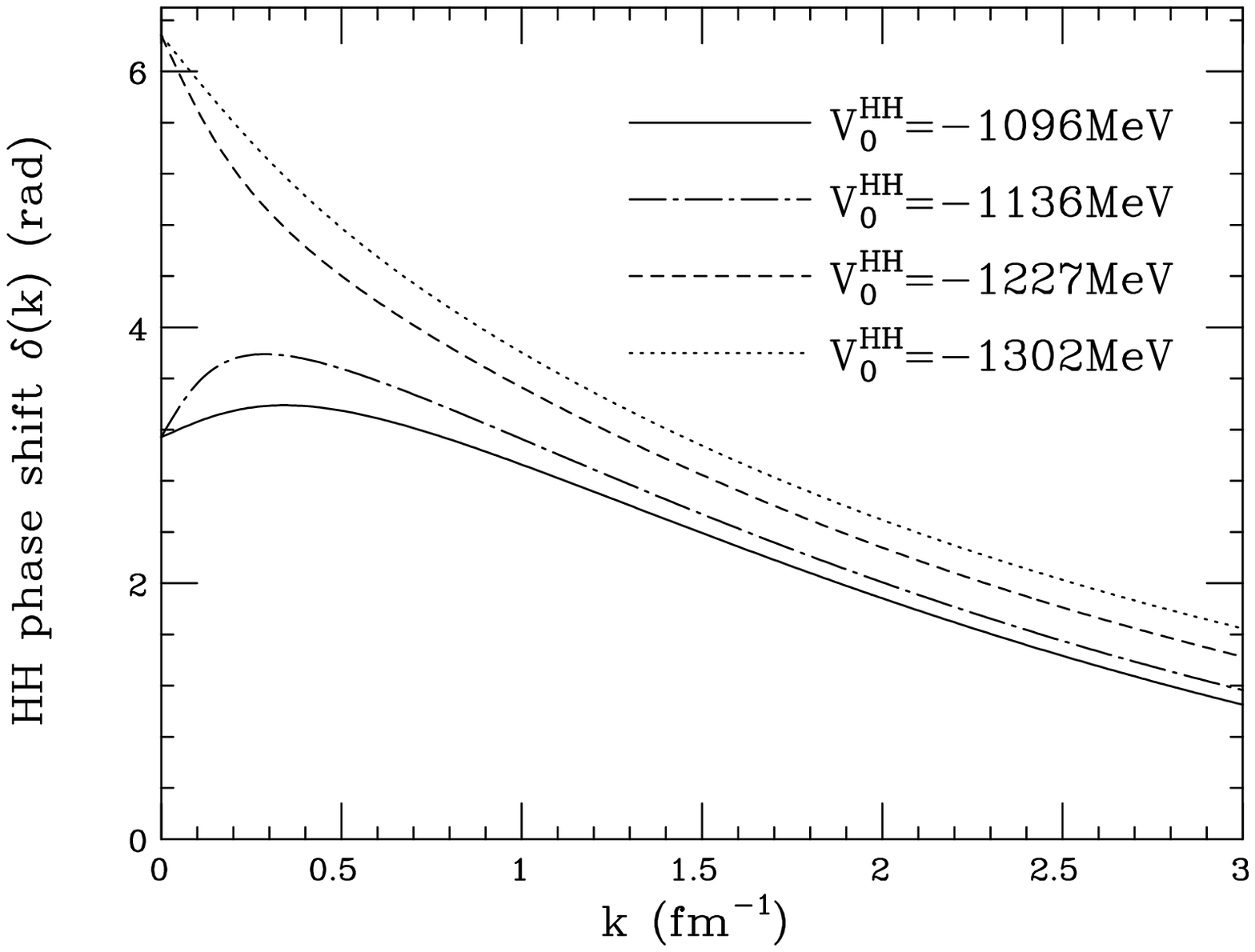}}
\caption{The HH scattering phase shift $\delta$ as a function of the 
relative wave number $k$ for several $V_0^{\rm HH}$. 
(Taken from Ref.~{\protect\citen{Sakai97}}.)}
\label{fig:HH-ps}
}
\end{figure}

In Fig.~\ref{fig:HH-BE} the binding energy of the HH system as a function of 
$V_0^{\rm HH}$ is shown.
The HH scattering phase shift is shown in Fig.~\ref{fig:HH-ps} as a function 
of the relative wave number $k$ for several $V_0^{\rm HH}$.
The following results are obtained from this calculation.

(i) The interaction between H-dibaryons has a strongly attractive nature, and
bears the deeply bound state.
The binding energy in the ground state is 100$\sim$170 MeV.
Even a second bound state appears for $V_0^{\rm HH} \lsim -1200$ MeV. 
(See Fig.\ref{fig:HH-BE}).

(ii) The main properties of the interaction between H-dibaryons can be 
characterized as a short-range repulsion due to Pauli blocking and the 
CMI, and a medium-range attraction due to 
flavor singlet scalar meson exchange (See Fig.\ref{fig:HH-ps}).
The two H-dibaryons are bound with a separation of about 0.8 $\sim$ 0.9 fm due 
to the strong repulsive core.
If we follow the discussion made in \S\ref{sec:NH}, the normalization 
kernel at zero separation is $\frac{25}{32}$, which indicates the effect of 
the Pauli blocking, and the CMI term bears a repulsive force as 
$\Theta_{\rm HH} - 2 \times \Theta_{\rm H} = 24 -2 \times (-24) = 72$.

It is worth while noting that a bound state of two H-dibaryons is also 
obtained in the Skyrme model\cite{Issinskii88} although 
the binding energy of the ``tetralambda" state ($E_{\rm B} = 15 \sim 20$ MeV)
is rather smaller than the value obtained in the QCM.
The attractive nature of the inter-H interaction is also implied 
in a study on H-dibaryon matter in the Skyrme model\cite{Sakai98}  
following the method developed by Manton {\it et al.}\cite{Manton86}

The implications of this result for the occurrence of H-matter are as follows.
In Tamagaki's discussion\cite{Tamagaki91} that there is a possibility of a 
phase transition from neutron matter to H-matter at a density which is 6$\sim$9 
times greater than the normal nuclear density $\rho_0$, the interaction 
between H-dibaryons is through a hard core potential and an attractive square 
well potential outside the core. 
The depth of the attractive potential was assumed to be so 
weak that it can be treated as a perturbation.
The depth of the square well potential has been determined so that 
the scattering length becomes zero for the first time,
when the strength of the attractive potential is gradually increased.
The pertinent strength parameter corresponding to the EMEP can 
be obtained from the scattering length calculated from S-matrix through 
the effective range theory. 
It is found to be $V_0^{\rm HH} = -638$ MeV. 
Thus the attraction used in Ref.~\citen{Tamagaki91} 
is much weaker than the one used in this QCM calculation, in which 
$V_0^{\rm HH}$ is typically $-1300 \sim -1100$ MeV.
If the attractive H--H potential is indeed as strong as the EMEP 
employed in this calculation, 
the critical transition density beyond which H-matter formation is 
energetically favorable may be appreciably lower, although a more 
quantitative estimate is not simple because of the inapplicability of 
perturbation theory under a strong attractive potential as in the present 
calculation.

Some comments on implications of the existence of the H-dibaryon to 
neutron star properties are in order.
In the framework of the Walecka model with the strength of the 
H-H interaction of the present calculation, it has been shown that 
H-matter is unstable against compression.\cite{Faessler97}
Therefore, if the central density of a neutron star exceeds the critical 
density for H-matter formation, the energetically favorable compression 
of H-matter could provide a scenario for the conversion of a neutron 
star into a strange quark star.\cite{Faessler97}

Using a relativistic mean field theory, it is studied how H-dibaryon 
condensate affects the equation of state and the properties of neutron 
stars.\cite{Glendenning98}
It is shown that, if the limiting neutron star mass is about the mass 
of the Hulse-Taylor pulsar (1.44$M_{\odot}$), a condensate of H-dibaryons 
with their mass in the vacuum about 2.2 GeV and a moderately attractive 
potential in the medium could not be ruled out.

\section*{Acknowledgements}
One of the authors (T.S.) is grateful for the Special COE grant of the 
Ministry of Education, Science and Culture of Japan.
The work is partly supported by Grant-in-Aid for Scientific 
Research on Priority Areas (Theoretical Study of Nuclei with Strangeness) 
by the Ministry of Education, Science, Sports and Culture.
The authors would like to thank A.~J. Buchmann and J. Mori for the 
collaboration on which this review is based.

%

\end{document}